\documentclass{article}
\usepackage[paper=a4paper,margin=1.2in]{geometry} % only for the title page

\usepackage[export]{adjustbox}
\usepackage{afterpage}
\usepackage{amssymb}
\usepackage{amsthm}
\usepackage{amsmath}
\usepackage{booktabs}       % used for toprule, midrule, etc.
\usepackage{chemformula}
\usepackage{comment}        % commenting
\usepackage{crimson}
\usepackage{draftfigure}
\usepackage{float}
\usepackage[hidelinks]{hyperref}
\usepackage[utf8]{inputenc}
\usepackage{multirow}
\usepackage{setspace}
\usepackage{soul}           % highlighting
\usepackage{subcaption}
\usepackage{tabularx}
\usepackage{xcolor}

\usepackage[
backend      = biber,
natbib       = true,
citestyle    = numeric,
bibstyle     = numeric,
date         = year,
url          = true,
urldate      = long,
isbn         = false,
doi          = true,
maxcitenames = 1,
maxbibnames  = 9,
uniquelist   = true,
sorting=none
]{biblatex}

\DeclareFieldFormat*{url}{}
\DeclareFieldFormat[online]{url}{\mkbibacro{URL}\addcolon\space\url{#1}}
\DeclareFieldFormat[report]{url}{\mkbibacro{URL}\addcolon\space\url{#1}}
\DeclareFieldFormat*{urldate}{}
\DeclareFieldFormat[online]{urldate}{\mkbibparens{\bibstring{urlseen}\space#1}}
\DeclareFieldFormat[article,misc,inbook,book,report,online]{title}{\mkbibquote{#1\isdot}} 

\renewbibmacro{in:}{}

\AtEveryBibitem{\clearlist{language}}

\newcommand{\footremember}[2]{%
    \footnote{#2}
    \newcounter{#1}
    \setcounter{#1}{\value{footnote}}%
}
\newcommand{\footrecall}[1]{%
    \footnotemark[\value{#1}]%
} 

\title{Power sector benefits of flexible heat pumps}

\author{%
  Alexander Roth\footremember{DIW}{German Institute for Economic Research (DIW Berlin), Mohrenstraße 58, 10117 Berlin, Germany}\footnote{Corresponding Author (\href{mailto:aroth@diw.de}{aroth@diw.de)}
}%
  \and Carlos Gaete-Morales\footrecall{DIW}%
  \and Dana Kirchem\footrecall{DIW}%
  \and Wolf-Peter Schill\footrecall{DIW} %
  }
\date{\today}

\begin{document}

% ------------------------------------------------------------------

\maketitle

% ------------------------------------------------------------------

\begin{abstract}

\noindent Heat pumps play a major role in decreasing fossil fuel use in heating. They increase electricity demand, but could also foster the system integration of variable renewable energy sources. We analyze three scenarios for expanding decentralized heat pumps in Germany by 2030, focusing on the role of buffer heat storage. Using an open-source power sector model, we assess costs, capacity investments, and emissions effects. We find that investments in solar photovoltaics can cost-effectively accompany the roll-out of heat pumps in case wind power expansion potentials are limited. Results further show that short-duration heat storage substantially reduces the need for firm capacity and battery storage. Larger heat storage sizes do not substantially change the results. Increasing the number of heat pumps from 1.7 to 10 million units could annually save around a quarter of Germany's overall natural gas consumption and around half of households' building-related \ch{CO2_{eq}} emissions.

\end{abstract}

% ------------------------------------------------------------------

\doublespacing

% ------------------------------------------------------------------
\section{Introduction}
% ------------------------------------------------------------------

In light of the climate crisis, heat pumps are regarded as a central technology to reduce greenhouse gas emissions in the heating sector \cite{iea_future_2022}. Heat pumps can displace traditional heating technologies such as oil- and gas-fired heating and mitigate greenhouse gas emissions, when powered with electricity from renewable energy sources (RES). In addition, in the European context, the Russian invasion of Ukraine has led to further political push, especially in Germany, to reduce the dependence on Russian natural gas imports. In Germany, natural gas is currently still the dominant source of residential heating. Therefore, the electrification of heating can be considered a critical measure to reduce the use of natural gas.

In Germany, policymakers are working to accelerate the implementation of decentralized heat pumps, with a declared target of 6 million heat pumps installed by 2030 \cite{roth_renewable_2023}. Given the current stock (2024) of around 1.7 million heat pumps, such a transition implies an increase in the electricity demand. So far, it is not yet fully understood how a larger heat pump stock affects the power sector in detail, considering that the electricity needs for mobility, hydrogen production, and other energy services will also rise. One common concern is that heat pumps could add to existing load peaks due to electricity load profiles coinciding with heat demand profiles, and thus increase the need for firm generation capacity or electricity storage. Therefore, the potential benefits of flexible heat pump operations are of central interest. For that reason, we explore the power sector effects of various heat pump rollout scenarios in Germany. In particular, we focus on different degrees of temporal flexibility in heat pump operations by varying the size of the heat storage assumed to be attached to heat pumps. To do so, we use the open-source capacity expansion model DIETER \cite{zerrahn2017review,schill2018results,gaete-morales_open_2021} to model the central European power sector for various scenarios of 2030.

%Literature
Previous studies have highlighted the important role of heat pumps in the decarbonization of the heating sector. A recent study shows that deploying heat pumps is one of the fastest strategies to reduce natural gas consumption in the German heating sector \cite{altermatt_replacing_2023}. Several studies investigate the potential of heat pumps to facilitate the integration of renewable energy sources in the power sector. For example, different analyses show that deploying additional heat pumps aligns well with additional investments into wind power plants \cite{ruhnau_heating_2020,chen_impact_2021}. Regarding the flexibility of heat pumps and optimal heat storage size, the picture is inconclusive. Investigating various heat storage sizes, one study finds that the optimal thermal energy capacity of heat pumps in Spain and the UK lies between 12 and 14~hours of maximum heat output \cite{lizana_national_2023}. A previous analysis of wind power deployment in Denmark finds that the flexible operation of heat pumps provides only moderate system benefits, and that even inflexible heat pumps enable a higher share of wind power energy \cite{hedegaard_influence_2013}. Heat pump flexibility can also be provided by the thermal inertia of buildings \cite{papaefthymiou_potential_2012}, which can help to integrate renewable energy \cite{rinaldi2022}. Another study points out that the power system cost savings from flexible electric heating with night storage in Germany are moderate because renewable availability patterns do not align well with heat demand profiles \cite{schill_flexible_2020}. The seasonal demand pattern disadvantages flexible electric heating compared to other sector coupling options without this seasonality, such as electric vehicles. This finding is also supported by another study \cite{kroger_electricity_2023} that identifies a larger potential for load shifting in electric vehicles than in heat pumps. Another study focuses on the role of flexible, large-scale, centralized heat pumps in district heating grids \cite{bernath_influence_2019}, finding a correlation between RES expansion and the choice of heating technologies. With higher deployment of RES, large heat pumps become more competitive. Other studies focus on the competition of flexibility provided by heat pumps with electricity storage units. In power systems with a share of renewable electricity of 80\% or higher, the flexible use of heat pumps reduces the investment needs for short-term electricity storage considerably \cite{hilpert_effects_2020}. The substitutional nature between pumped hydro storage and thermal storage is also highlighted in the literature \cite{ruhnau_heating_2020}.

Our paper adds to the existing body of literature by investigating the power sector effects of decentralized heat pumps in detail, specifically accounting for different amounts of temporal flexibility facilitated via heat storage. In our analysis, we use an open-source capacity expansion model that considers the hourly variability of renewable electricity generation and heat demand over an entire year. It also accounts for additional electric load related to electric vehicles and the production of green hydrogen. To the best of our knowledge, such an analysis has not been done so far. We investigate how different rollout speeds of heat pumps in Germany, specifically in combination with different heat storage capacities, impact the optimal capacity investment and dispatch decisions in the power system. In addition, we also provide an ex-post calculation to measure the associated natural gas usage, cost, and emission savings. To check the robustness of our results, we further carry out numerous sensitivity analyses with alternative assumptions on relevant input parameters such as renewable availability, including an extended drought period, different natural gas prices, and a German coal phase-out.

We show that an ambitious roll-out of decentralized heat pumps can be accommodated in the German power sector at relatively low costs. If wind power expansion potentials are limited, heat pumps can be accompanied by an expansion of solar photovoltaic (PV) capacities in combination with moderate additions of electricity storage and other firm generation technologies, while also leveraging flexibility potentials of the European interconnection. Short-duration heat storage helps to integrate renewable electricity efficiently and reduce peak loads, hence diminishing the need for additional generation and storage capacities. We find a strongly diminishing additional system value of longer-duration heat storage. Finally, we assess the impact of the heat pump roll-out on German natural gas consumption, \ch{CO2_{eq}} emissions, and costs. We conclude that an ambitious roll-out can contribute significantly to decreasing Germany's gas consumption, while reducing emissions and overall system costs.

% ------------------------------------------------------------------
\section{Results} \label{sec:base-results}
% ------------------------------------------------------------------

\subsection{Heat pump rollout triggers investments in electricity generation and storage}

In the following, we present the power sector effects of three different roll-out speeds of heat pumps (\textit{slow}, \textit{government}, and \textit{fast}). In these, the number of heat pumps increases from 1.7 million units (in the \textit{reference} scenario) to 3.0, 6.0, and 10.0 million respectively (see Table~\ref{tab:hp_stocks}). We also discuss how the size of short-duration buffer heat storage attached to decentralized heat pumps affects the results. All these are part of our set of \textit{baseline} scenarios. In these, we assume expansion limits of 115~gigawatt (GW) for onshore wind power and 30~GW for offshore wind power, no regulated phase-out of coal-fired power plants, and a natural gas price of 50~Euro per MWh. 

% case with 0 heat storage
Expanding the stock of heat pumps requires additional investments into electricity generation infrastructure. Looking first at inflexible heat pumps, we find that in the \textit{reference} rollout, the cost-optimal capacity mix for reaching 80\% renewable energy in Germany entails 115 GW onshore wind, 26 GW offshore wind, and 98 GW of solar PV (Figure \ref{fig:investments_base}, panel A). Further, 10~GW of hard~coal and 47~GW of gas-fired power plants (sum of open and closed cycle gas turbines) are present. A rollout beyond the reference scenario requires higher generation capacity additions. The total additional generation capacities are around 5~GW and 20~GW (panel B) in the scenarios \textit{slow} and \textit{government}. For the \textit{slow} rollout, 2~GW of offshore wind power and less than 1~GW of solar PV capacities are added, while around 3~GW combined of gas-fired power plants and lithium-ion batteries are added. For the \textit{government} rollout, these numbers increase to 7~GW of additional gas-fired power plants and 8~GW of additional solar PV, driven partly by the fact that offshore wind capacities have reached their upper bounds and can only be expanded by an additional 4~GW. In the scenario \textit{fast} with the highest rollout of 10~million heat pumps, 57~GW of solar PV capacity is added. In parallel to this large expansion of solar PV capacities, firm capacities in the form of gas-fired power plants increase by 18~GW to ensure the coverage of peak loads, while also 9~GW of lithium-ion battery storage are added. The optimal storage energy capacity of batteries increases by 3, 8, and 49~gigawatt-hours~(GWh) in the three respective rollout scenarios (Figure~\ref{fig:investments_full}).

% varying heat storage of HPs
\subsection{Heat storage reduces capacity needs for electricity generation and storage}\label{subsec:base-results-capacities}

\begin{figure}[H]
    \centering
    \caption{Capacity investments under baseline assumptions with different heat pump rollouts and heat storage sizes}
    \includegraphics[width=\textwidth]{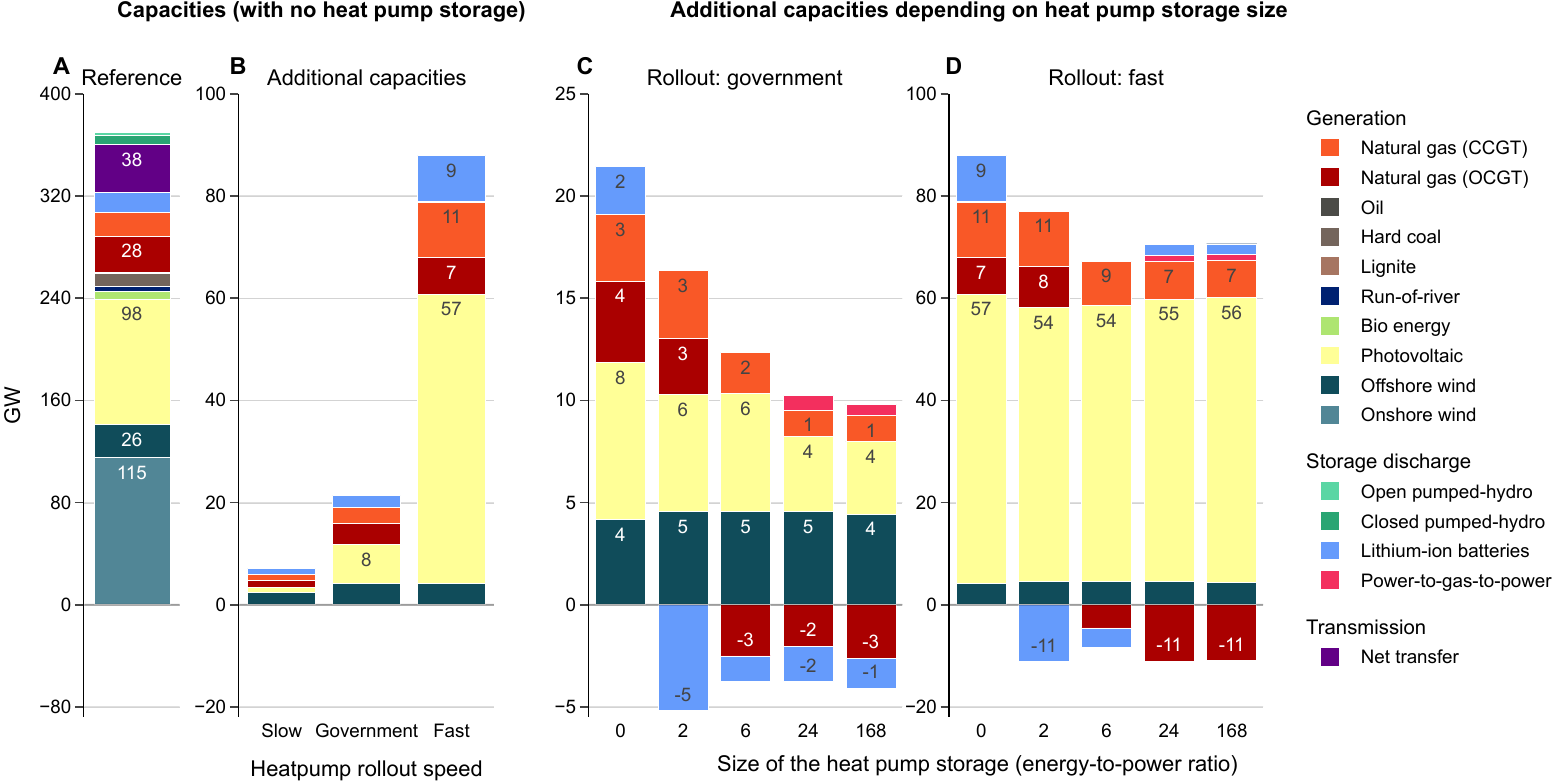}
        \begin{minipage}[c]{\textwidth}
        \bigskip
        \footnotesize
        \textbf{A} Optimal capacity investments in the \textit{reference} scenario. \textbf{B} Changes induced by the rollout of heat pumps in the case of inflexible heat pumps. \textbf{C-D} Capacity changes to the respective references for different heat storage sizes in the \textit{government} rollout (\textbf{C}) and \textit{fast} rollout (\textbf{D}). The changes shown in \textbf{C} \& \textbf{D} are to their respective references with different heat storage sizes. Reference results (\textbf{A}) are almost identically for different heat pump storage sizes. For better visibility, only one \textit{reference} rollout is shown. Please note the different y-axis ranges of the different panels. The complete set of results, including those for storage energy, is shown in Figure~\ref{fig:investments_full}.
        \end{minipage}
    
    \label{fig:investments_base}
\end{figure}

Equipping heat pumps with heat storage reduces the need for additional electricity generation and storage capacities (Figure~\ref{fig:investments_base}, panels~C~\&~D). Introducing heat storage with an energy-to-power~(E/P) ratio of 2~hours reduces the need for additional solar PV capacities (e.g., 6~GW instead of an additional 8~GW in the \textit{government} rollout) compared to the \textit{reference}. In addition, the need for battery storage is reduced by around 7~GW compared to the case without heat storage, and even by 2~GW compared to the \textit{reference}. This effect can be explained as lithium-ion batteries and heat storage of heat pumps are both short-duration storage technologies and, therefore, serve as complements, especially when taking up daily PV surplus generation peaks. Expanding the heat storage of heat pumps beyond an E/P~ratio of 2~hours, even lower additional solar~PV capacities are needed, but especially additional capacities of gas-fired power plants are reduced. The capacity-reducing effect of the buffer storage of heat pumps can also be seen in the decreasing left-hand side of the residual load duration curves (Figure~\ref{fig:rldc}). If combined with heat storage larger than 6~hours in the \textit{government} rollout (panel~C) or larger than 24~hours in the \textit{fast} rollout (panel~D), the introduction of heat pumps even reduces the overall need for gas-fired power plants as compared to the reference. Qualitative results are largely similar between the \textit{government} and \textit{fast} rollout.

We find a substitution between lithium-ion batteries and heat storage not only for storage power but also for storage energy capacities (Figure~\ref{fig:investments_full}). While the deployment of heat pumps leads to additional lithium-ion energy capacities in all three rollouts, the introduction of a 2-hour heat energy storage does not only reduce the additional need for storage energy capacities but even turns it negative: hence the introduction of heat pumps in combination with 2-hour heat storage is reducing the overall need for lithium-ion energy capacities. For larger heat storage capacities of 24 and 168 hours, this absolute reversal cannot be detected, yet additional energy capacities are still considerably lower than in the case of inflexible heat pumps.

Due to the interconnection with its neighboring countries, the heat pump expansion in Germany could be partly supported by non-German generation and storage capacities. To avoid unintended support of German heat pumps by foreign power plants, we also co-optimize the power plant portfolios of neighboring countries, assuming an upper limit for fossil-fuel power plants outside Germany. Therefore, we ensure that German heat pumps in the model do not unduly benefit from an oversized exogenous power plant fleet outside Germany. Figure \ref{fig:investment_other_countries} confirms that aggregated generation capacities in all countries except Germany barely change after the introduction of heat pumps in Germany.

\subsection{Heat storage helps to integrate renewable electricity}\label{subsec:dispatch}

\begin{figure}[H]
    \centering
    \caption{Yearly electricity generation by source under baseline assumptions}
    \includegraphics[width=\textwidth]{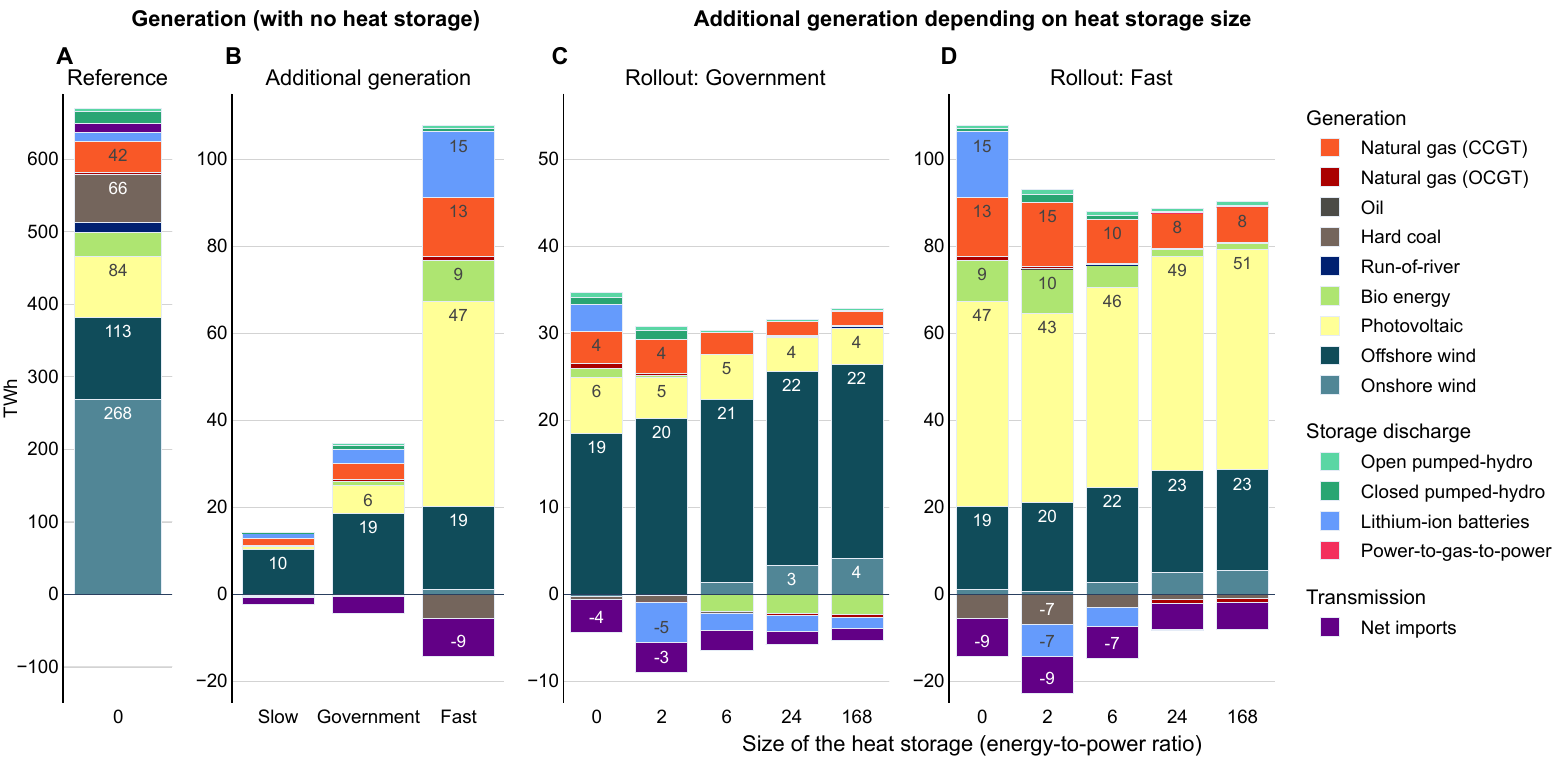}
    \begin{minipage}[c]{\textwidth}
    \bigskip
    \footnotesize
    \textbf{A} Optimal dispatch in the reference scenario. \textbf{B} Changes induced by the rollout of heat pumps in the case of inflexible heat pumps. \textbf{C-D} Changes in dispatch to the respective \textit{reference} scenarios for different heat storage sizes in the \textit{government} rollout (\textbf{C}) and \textit{fast} rollout (\textbf{D}). The changes shown in \textbf{C} \& \textbf{D} are to their respective \textit{reference} scenarios with different heat storage sizes. The results for the different \textit{reference} scenarios (\textbf{A}) are almost identically for different heat pump storage sizes. For better visibility, only one \textit{reference} rollout is shown. Please note the different y-axis ranges of the different panels. The complete set of results is shown in Figure~\ref{fig:dispatch_full}.
        \end{minipage}
    \label{fig:dispatch_base}
\end{figure}

The rollout of heat pumps affects optimal generation and storage discharge (Figure~\ref{fig:dispatch_base}) similarly to optimal capacities. The additional electricity needed for heat pumps is primarily generated by offshore wind power and solar PV. The latter plays a more important role in the \textit{fast} rollout due to the upper capacity bound of offshore wind power. As the profiles of solar PV generation and heat pump load only align to some extent, the expansion of heat pumps triggers additional generation by gas-fired power plants. Battery storage is also used more in the case of inflexible heat pumps, but less if heat storage is available. If no heat storage is available, the rollout of heat pumps is also accompanied by additional generation from bio energy, which is another flexible generation technology, but comes with relatively high variable costs. Accordingly, bio energy use decreases (or turns even negative in the \textit{government} rollout) if heat storage is available. Net imports of electricity slightly decrease with the rollout of heat pumps, especially when they do not come with heat storage, i.e., are operated inflexibly. That is due to more exports of renewable energy surpluses triggered by additional renewable energy capacities needed for the additional heat pumps. Figure~\ref{fig:dispatch_full} shows the dispatch results of all scenarios.

\begin{figure}[H]
    \centering
    \caption{Exemplary weeks of electricity generation, heat pump operation, and wholesale prices}%
    \includegraphics[width=\textwidth]{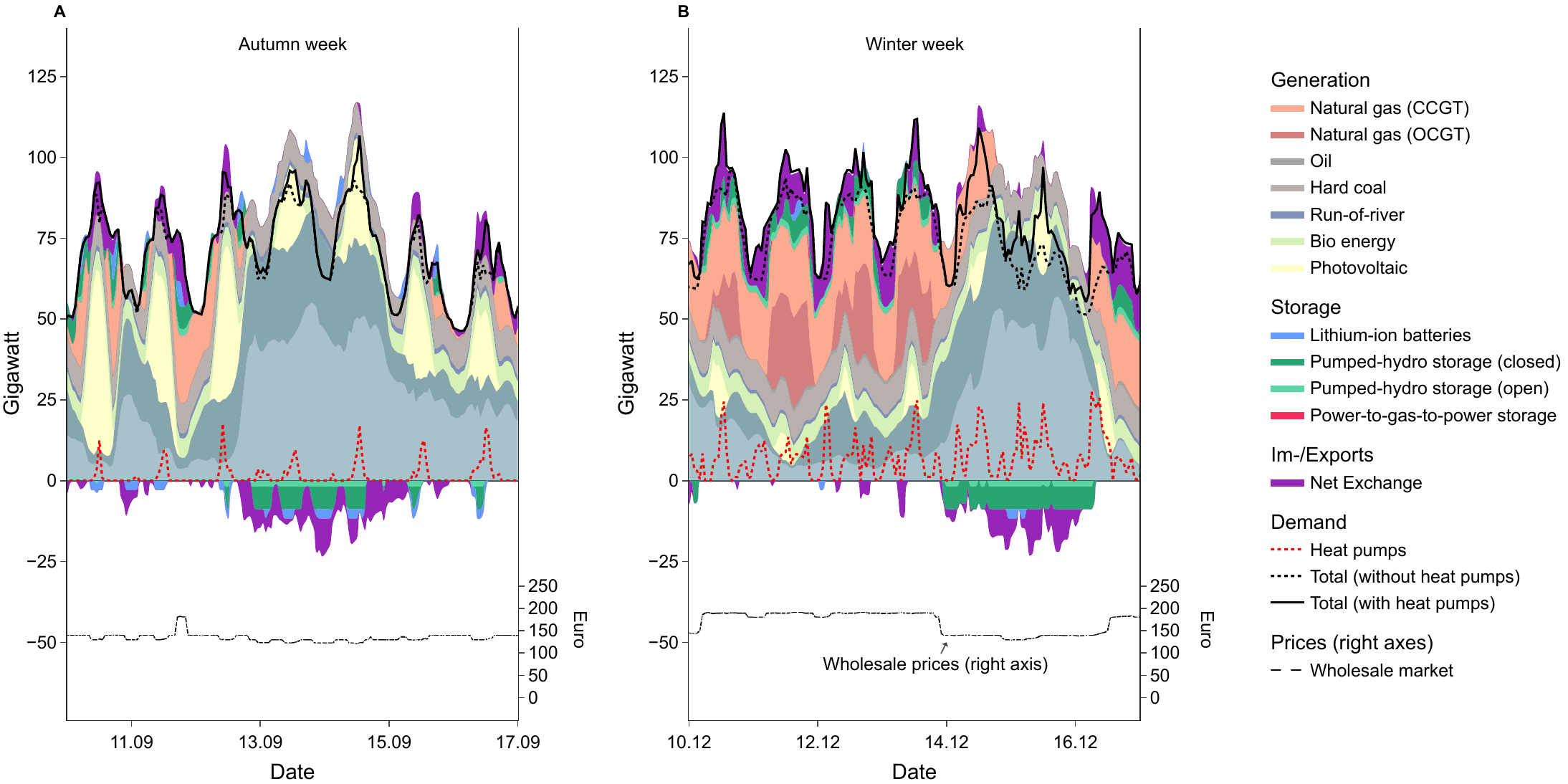}
    \begin{minipage}[c]{\textwidth}
    \bigskip
    \footnotesize Two exemplary weeks are shown for the \textit{government} rollout of heat pumps with heat storage of two hours: \textbf{A} autumn week and \textbf{B} winter week.   
    \end{minipage}
    \label{fig:exemplary-week}
\end{figure}

While the capacity and dispatch results already show that heat storage can help integrate renewable energy into the energy system, this effect is highlighted in Figure \ref{fig:exemplary-week}. It provides an illustration of hourly electricity generation and heat pump operation in combination with additional heat pumps. The figure depicts two exemplary weeks under \textit{baseline} assumptions with a \textit{government} rollout and 2~hours of heat storage. The diurnal fluctuations of solar PV generation are clearly visible, especially in the autumn week. In contrast, wind power generation has less regular yet longer variability patterns. In hours of low wind and solar PV generation, gas-fired power plants and imports cover the remaining residual load. Even with only 2~hours of heat storage capacity, heat pumps can align a substantial part of their electricity consumption with PV peak generation periods. This indicates that even small heat storage capacities already improve the integration of heat pumps into the system. Hours of electricity exports, storage charging, and heat pump use often coincide, which are also hours with relatively low prices. Conversely, heat pumps largely avoid drawing electricity from the grid during hours when imports take place, which often coincides with hours of low renewable generation and relatively high prices.

\begin{figure}[H]
\centering
    \caption{Heat output, heat pump electric load with different storage sizes, and residual load}
    \includegraphics[width=\textwidth]{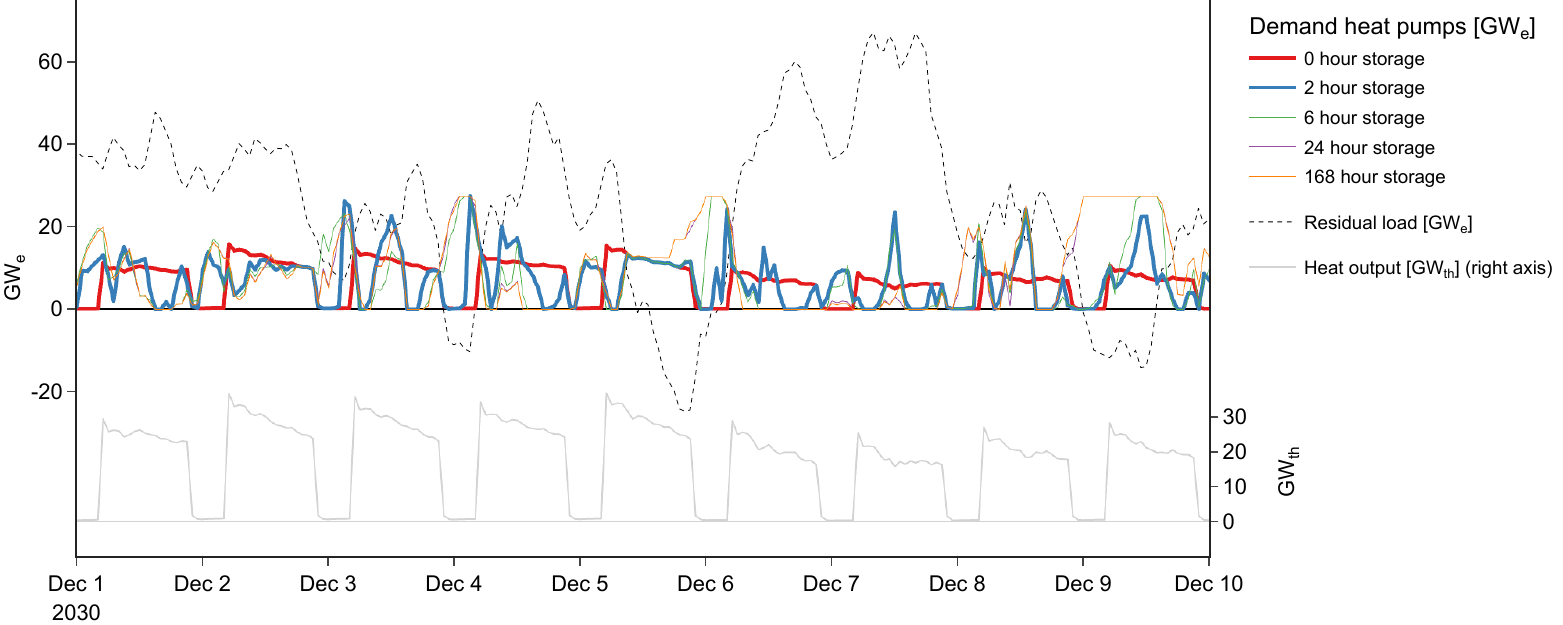}
    \begin{minipage}[c]{\textwidth}
    \bigskip
    \footnotesize
    Residual load, the heat output of heat pumps, and their electricity demand for different heat storage sizes in the \textit{baseline} setting with a \textit{government} rollout are shown.
    \end{minipage}
    \label{fig:hp-draw}
\end{figure}

In our model, heat pumps are operated in a way that minimizes system cost, which can be interpreted as if they are following wholesale market price signals. The presence of heat storage enables and increases their potential to do so. As visible in Figure~\ref{fig:hp-draw}, there is a strong alignment of heat pump electricity intake and relatively low residual load levels when heat pumps are equipped with heat storage. Heat pumps with no heat storage are inflexible electricity consumers, which directly follow the hourly heat demand profile. In Figure~\ref{fig:hp-draw} that is visible by the parallel movement of heat output (gray line, right y-axis) and the electricity demand in case of no heat storage (red line, left y-axis). This changes when heat storage is added. Even small two-hour heat storage makes heat pumps sufficiently flexible that they can adjust their demand to the overall system conditions to a considerable extent. If heat storage is expanded further, heat output and electricity intake are even less correlated. For the \textit{fast} rollout scenario, results are more pronounced but qualitatively similar (Figure~\ref{fig:hp-draw-fast}).

\begin{figure}[H]
    \centering
    \caption{Heatmap of the electricity demand of heat pumps for different heat storage sizes}
    \includegraphics[width=\textwidth]{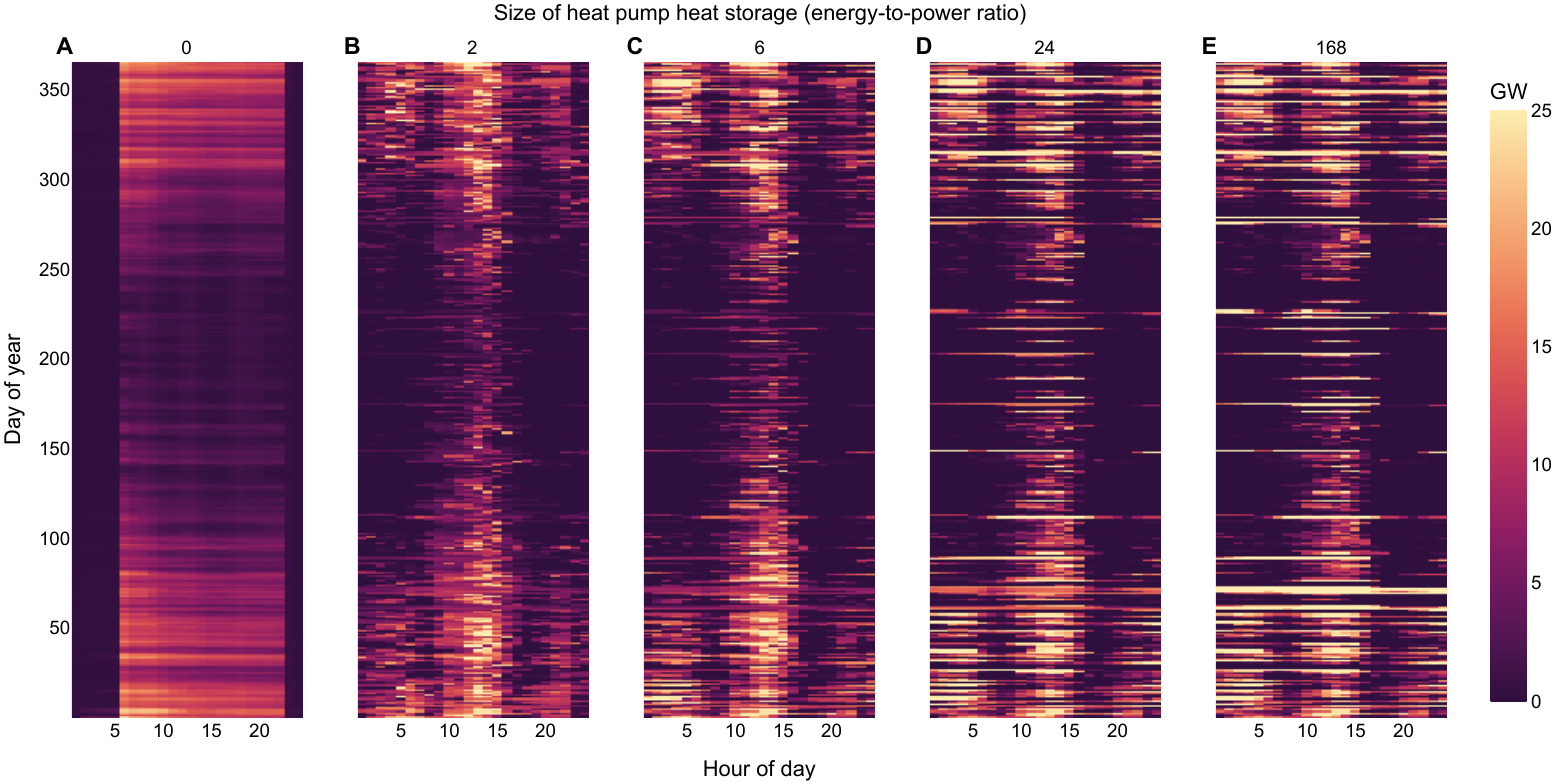}
        \begin{minipage}[c]{\textwidth}
        \bigskip
        \footnotesize \textbf{A}-\textbf{E} show the electricity demand of heat pumps for different heat storage sizes. Values are from the \textit{baseline} scenarios with \textit{government} rollout.
        \end{minipage}
    \label{fig:hp load heatmap gov}
\end{figure}

The shifting of electric loads through heat storage is also depicted in Figure~\ref{fig:hp load heatmap gov}. The figure shows the electricity demand of heat pumps depending on the size of the heat storage over the course of an entire year. Panel~A in Figure~\ref{fig:hp load heatmap gov} shows the electricity demand of heat pumps without heat storage, which mirrors the heat demand (also shown in Figure~\ref{fig:heatmap heatdemand}). We see that in the winter months, hence on the bottom and top of every panel, demand is higher than in the summer months. In all days of the year, heat demand during the first hours of the day is assumed to be zero. Moving from panel~A to the other panels B-E on the right, we see that the electric load patterns change when heat pumps are operated in a more flexible manner. Already with heat storage of 2~hours, heat pumps are used to integrate excess solar energy from the middle of the day. Furthermore, due to the flexibility enabled by heat storage, heat pumps can draw electricity in hours of no heat demand, such as at night, and hence smooth heat consumption peaks in morning hours. For larger heat storage sizes, such as 24 and 168~hours, the electric load of heat pumps increasingly resembles the charging of a longer-duration storage asset, sometimes consuming excess electricity for extended periods and avoiding consumption later. While this mode of operation may not be realistic for small, decentralized heat pumps due to limited potential for low-cost heat storage installations, such operational patterns appear more plausible for centralized heating solutions with larger and lower-cost heat storage options.

\subsection{Heat storage reduces electricity sector costs} \label{subsec:system-costs}

\begin{figure}[H]
    \centering
    \caption{Additional electricity sector costs and break-even investment costs of heat storage}
    \includegraphics[width=\textwidth]{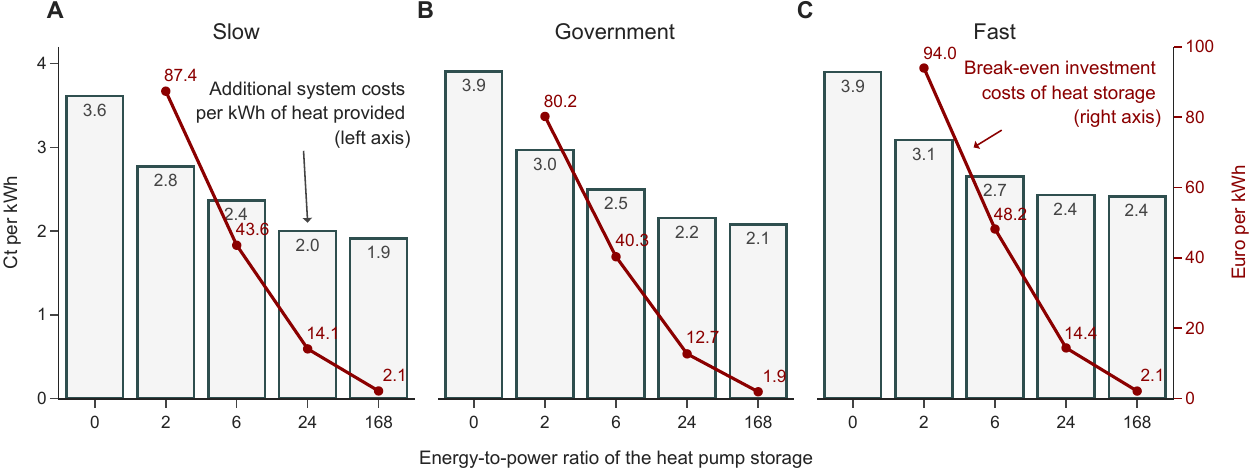}
    \begin{minipage}[c]{\textwidth}
        \bigskip
        \footnotesize
        For different rollout scenarios (\textbf{A-C}), the gray bars depict the additional costs in Euro-cent/kilowatt-hours$_{\text{th}}$ of additional heating energy provided for different rollout scenarios and heat storage sizes (compared to the respective reference, left y-axis). The red lines show the break-even investment costs in Euro/kilowatt-hours$_{\text{th}}$ of heat storage (compared to the respective scenario without heat storage, right y-axis).
        \end{minipage}
    \label{fig:aschp_base}
\end{figure}

Our analysis focuses on additional electricity sector costs caused by the heat pump expansion. We relate these costs to the additional heating energy provided (Figure~\ref{fig:aschp_base}). More heat pumps lead to additional costs for the electricity sector due to additional investment into generation and storage capacities and higher variable costs. We find that electricity sector costs increase by around four~Euro-cent/kilowatt-hour (ct/kWh) of additional heating energy provided in the \textit{government} rollout scenario with inflexible heat pumps. 

This cost effect decreases with larger heat storage sizes. The relative decline in additional costs is the largest when moving from no storage to a 2-hour~storage. With larger heat storage sizes, the decreases become smaller and are minimal between a day (24~hours) and a week of heat storage (168~hours). This means that the additional value of long-duration storage compared to shorter-duration storage is relatively small in the modeled setting with an 80\% renewable share in Germany. In other words, the marginal electricity sector cost savings decrease with larger heat storage. The cost effects shown in Figure~\ref{fig:aschp_base} do not include the installation costs of heat pumps and heat storage, but only the costs related to the electricity sector, such as investment and operational expenses of generation and electricity storage capacities. Therefore, we can interpret these figures as opportunity costs of heat storage. 

We calculate the break-even overnight investment costs of heat storage that would be required to deliver overall system cost savings. We do so by relating the power sector cost differences between scenarios with different heat storage sizes to the respective storage capacity and deriving annualized overnight investment costs. For the latter, we assume that heat storage installations have a lifetime of 20~years, face an interest rate of 4~percent, and do not incur any variable or fixed operation and maintenance costs. As the marginal power sector cost savings decrease with larger heat storage capacities, the specific break-even costs of heat storage decrease even faster. For example, specific heat storage investment costs have to be below around 80~Euro/\ch{kWh_{th}} in the \textit{government} rollout with 2 hours of heat storage. This break-even cost decreases to 40~Euro/\ch{kWh_{th}} in the case with 6 hours of storage and sharply declines to only about 2~Euro/\ch{kWh_{th}} for a heat storage size of one week. In other words, long-duration heat storage would have to be very cheap. Results are qualitatively similar for the \textit{slow} and \textit{government} rollouts. In the latter, break-even costs are slightly higher at around 94~Euro/\ch{kWh_{th}} in the 2-hour case, as the additional PV generation in this scenario increases the value of temporal flexibility. While the real-world costs of short-duration heat storage technologies are likely well below the range of 2-hour break-even costs determined here, we consider it implausible that storage sizes of a day or more could be realized in combination with decentralized heat pumps at the required costs. Low-cost, long-duration heat storage appears much more plausible in combination with district heating networks and large-scale, centralized heat pumps, which is beyond the scope of our analysis.

% ----------
\subsection{Qualitative results also hold in sensitivity analyses} \label{sec:sensitivity}
% ----------

In addition to our \textit{baseline} scenario in which we vary the rollout speed of heat pumps and heat storage duration, we conduct several sensitivity analyses. Those help us judge how strongly our results hinge on certain fundamental model assumptions. We alter the assumptions on the upper bounds of wind power expansion, the level of natural gas prices, and the possibility of electricity generation from coal. We also introduce a week of a stylized ``renewable energy drought'' with no wind and solar availability. Table~\ref{tab:sensitivity scenarios} provides an overview of all sensitivity analyses. 

In the following, we summarize the principal takeaways of these sensitivity analyses. An extensive description and discussion of the results of the sensitivity analyses, including additional figures, are found in the Supplementary Information (Section~\ref{sec:sensitivity-appendix}).

In the \textit{baseline} scenarios, we assume an upper limit for on- and offshore wind power capacity expansion in Germany of 115~GW and 30~GW, respectively. This appears to be realistic and policy-relevant from a 2030 perspective. Considering real-world constraints related to regulation, land availability, and public acceptance, unbounded wind power capacity expansion seems implausible. Still, removing these limits (scenario \textit{no wind cap}), generates complementary insights into a less constrained equilibrium setting. Results show an increase in wind onshore capacities at the expense of offshore wind and a slight reduction in PV capacity. This leads to higher onshore wind generation and less offshore wind dispatch (Figure~\ref{fig:dispatch_sens}). The seasonality of heating demand aligns well with wind power and additional system costs due to heating decrease slightly compared to the \textit{baseline}.

Scenarios with higher gas prices (\textit{gas100} and \textit{gas150}) show fewer gas-fired power plants and more solar PV capacities in the \textit{reference} and different rollout scenarios. Especially in \textit{gas150}, additional capacities are mostly solar PV as offshore wind is already at its limit in the \textit{reference}. In parallel, dispatch sees reduced gas-fired generation both in the \textit{reference} and the rollout scenario (Figure~\ref{fig:dispatch_sens}). The system costs per heating unit increase considerably due to more expensive natural gas (Figure~\ref{fig:aschp_sens_si}). 

Without coal-fired power plants (\textit{coal phase-out}), gas-fired generation and electricity imports increase in the \textit{reference} rollout. Yet, the dispatch effects of a heat pump rollout do not substantially differ from the \textit{baseline} setting. Combining coal phase-out with higher gas prices leads to system effects similar to the scenarios \textit{gas100} and \textit{gas150}.

As the share of variable renewable energy increases, the security of supply during prolonged periods with low renewable energy supply becomes an increasing concern \cite{raynaud2018energy,kittel2024measuring}. Therefore, we assess how a week of a severe renewable energy drought in Europe would affect our results. To simulate an extreme case of such a week, we artificially set wind and solar PV capacity factors to zero in all modeled countries during one winter week. Such a week-long renewable energy drought (scenario \textit{RE drought}) requires substantially more firm capacity, which is provided by gas-fired power plants in the \textit{reference} (Figure~\ref{fig:investments_sens_si}). The effects of a heat pump rollout on electricity generation and storage technologies is also higher than in the \textit{baseline}, and we see higher cost increases. Yet, the overall yearly dispatch in the reference, and also the dispatch effects of a heat pump rollout, hardly change compared to the \textit{baseline}. Similarly, heat pumps still provide power system flexibility by aligning their electricity demand to the residual load if equipped with heat storage (Figure~\ref{fig:hp-draw-fast-badweek}).

Overall, our sensitivity analyses indicate that the key insights and results from the \textit{baseline} scenarios hold true under varying assumptions. The addition of heat pumps, depending on the rollout speed, requires additional investments in wind offshore, solar PV, gas-fired plants, and short-duration storage to meet renewable energy constraints. Unlimited wind power expansion offers some benefits, but overall cost reductions are modest.

% ----------
\subsection{An ambitious rollout of heat pumps leads to large savings of overall system cost, natural gas use, and carbon emissions} \label{sec:gas and emission savings}
% ----------

Based on the power sector optimization results, we also examine the effects of different rollout speeds of heat pumps on natural gas usage and carbon emissions. We compare the \textit{reference} rollout of 1.7~million heat pumps with 4.3~million additional heat pumps in the \textit{government} rollout scenario and 8.3 million additional heat pumps in the \textit{fast} rollout scenario. The underlying assumptions for the calculation of gas and emission savings are stated in Table~\ref{tab:gas_parameters}. Table~\ref{tab:gas_savings} summarizes the results. 

Under the assumption that each heat pump replaces one gas boiler with a thermal efficiency of 0.9\footnote{A thermal efficiency of 0.9 means that 1~kWh of natural gas will be transformed to 0.9~kWh of heat.}, additional heat pumps displace around 76~terawatt-hours\textsubscript{th} (TWh) of natural gas in case of a \textit{government} rollout and 224~TWh\textsubscript{th} with a \textit{fast} rollout (Table~\ref{tab:gas_savings} and Table~\ref{tab:savings_fullresults}), compared to the \textit{reference} rollout. At the same time, natural gas usage for electricity generation slightly increases in both scenarios, but this is by far overcompensated by natural gas savings in the heating sector, leading to total savings of around 206~TWh\textsubscript{th} in the \textit{fast} rollout compared to the \textit{reference} rollout. For the more moderate \textit{government} rollout with lower gas prices, overall natural gas savings still amount to around 73~TWh. To put these numbers into perspective, 73~(206)~TWh of natural gas corresponds to around 9~(24)~percent of Germany's overall natural gas consumption in 2022, or around a fifth (three-fifths) of private and commercial natural gas demand. In the scenarios with higher natural gas prices of 100~Euro or 150~Euro~per~ MWh\textsubscript{th}, we find largely similar effects on overall natural gas usage.

% german gas consumption in 2022
% total: 847.470 GWh
% households and commercials: 350.751 GWH
% source: https://www.bundesnetzagentur.de/DE/Gasversorgung/a_Gasversorgung_2022/start.html

We observe an increase in overall system cost savings with an increasing number of heat pumps and increasing gas prices. In this calculation, overall system cost effects include the increase in power sector costs due to higher electricity demand, the total annualized overnight investment costs of the additional heat pumps, the savings in natural gas expenditures, savings of \ch{CO2} emission cost of gas heaters, as well as investment costs of replaced natural gas boilers. As the investment- and installation costs of heat pumps might even fall below our cited values due to technical progress, our costs saving numbers can be interpreted as a lower bound. Overall system cost savings are between 2.0 and 6.7~billion~Euro in the \textit{government} or \textit{fast} rollout scenarios for a conservative natural gas price assumption of 50~Euro~per~MWh. Savings increase substantially with higher gas prices, up to 27.1~billion~Euro per year in the \textit{fast} rollout scenario with a gas price of 150~Euro~per~MWh.

Our overall system cost calculations depend on the assumption that every new heat pump substitutes a new gas boiler, which would otherwise have to be installed. We do not consider retiring existing gas boilers before the end of their lifetime. The extent to which an accelerated heat pump rollout would lead to a replacement of existing gas boilers, which have not yet reached the end of their lifetime, is unclear due to a lack of data on the age of existing gas boilers in the buildings modeled here. We calculate a counterfactual extreme case that assumes that the gas boilers replaced by heat pumps could have been used for another 20 years, which means we do not consider their investment costs in the calculation. This leads to smaller, but still positive, overall system cost savings below between 0.8~billion~Euro in the \textit{government} rollout and 4.6~billion~Euro in the \textit{fast} rollout. 

The reduced consumption of natural gas correspondingly leads to lower greenhouse gas emissions. In a \textit{fast} rollout scenario of heat pumps, emission savings in the range of 40-42~million~tons~\ch{CO2_{eq}} can be expected under different gas price assumptions. This corresponds to around 51\% of German households' carbon emissions from buildings in 2022. For the \textit{government} rollout, we can expect emission savings of around 14~million~tons~\ch{CO2_{eq}} (18\% of households' carbon emissions). Hence, an ambitious heat pump rollout, as described in this paper, could make a major contribution to Germany's carbon emission reductions. A further expansion of heat pumps beyond 2030 would lead to even higher reductions in carbon emissions. Note that the assumed emission factor of natural gas of 0.2~t\ch{CO2_{eq}}/\ch{MW_{th}} appears to be a conservative estimate, as it does not take methane leakage within the natural gas supply chain into consideration. Thus, the emission savings from switching to heat pumps presented here can be considered a lower bound and might be substantially higher because of methane leakage. 

% https://view.officeapps.live.com/op/view.aspx?src=https%3A%2F%2Fwww.umweltbundesamt.de%2Fsites%2Fdefault%2Ffiles%2Fmedien%2F361%2Fdokumente%2F2023_03_15_em_entwicklung_in_d_ksg-sektoren_pm.xlsx&wdOrigin=BROWSELINK

% 80,3 in 2022, Gebäude Haushalte

% ----------------------------------
\section{Conclusion}
% ----------------------------------

%Summary baseline results
As heat pumps are considered a key technology in the heating transition, their potential future impact on the electricity sector is of interest. We determine the effects of different rollout paths of decentralized heat pumps in Germany, combined with heat storage of different sizes, on the power sector. Under \textit{baseline} assumptions, we find that the expansion of the German heat pump stock from 1.7 to 10~million would require additional investments of around 54-57~GW of solar PV capacity in a least-cost solution, depending on how much heat storage is available. These results are driven by the assumption that the additional electricity consumption of heat pumps has to be covered by additional renewable electricity on an annual basis and that the expansion of wind power is limited to 115~GW (onshore) and 30~GW (offshore), respectively. For a slower rollout speed, which still achieves the German government's target of 6 million heat pumps by 2030, additional PV capacities of around 4-8~GW are needed.

Our results suggest a moderate need for additional firm capacity in the form of gas-fired power plants and lithium-ion batteries in most rollout scenarios, particularly in the \textit{fast} rollout scenario. More flexible heat pump operations facilitated by short-duration heat storage can partially relieve these additional capacity needs. While this is in line with several previous studies \citep{baeten2017,hilpert_effects_2020, ruhnau_heating_2020}, other research concluded that larger thermal heat storage of 12 to 24 hours would be desirable in Spain and the United Kingdom \citep{lizana_national_2023}. Our findings further corroborate previous research which found that heat pump deployment may increase optimal PV capacities \cite{rinaldi2022}. The European interconnection also helps to integrate heat pumps into the power sector and to limit additional capacity needs. This is in line with a previous analysis \cite{bernath_influence_2019}, which also highlights the importance of interconnection, as large-scale heat pumps become more competitive if they can make use of renewable surpluses in other countries.

Already small buffer heat storage with an energy capacity of 2~hours enables heat pumps to better align electricity consumption with the residual load. This results in power system cost savings of up to 0.9~ct/kWh of provided heat (or around 20\%) compared to a case with inflexible heat pumps. Costs further decrease with increasing heat storage, yet the marginal cost savings strongly decline with larger heat storage size. This hints at the fact that heat storage mainly serves to smooth daily renewable energy fluctuations. For 2-hour storage, it appears plausible that the costs of installing heat storage remain below the power sector benefits determined here. In contrast, long-duration heat storage would have to be very cheap to break even, which appears more plausible for large-scale thermal storage in district heating systems. 

%Summary sensitivity results
Sensitivity analyses show that results are generally robust against changes in key scenario assumptions. Assuming unconstrained expansion potentials for wind power substantially reduces solar PV capacity deployment since wind energy aligns better with heat demand \cite{ruhnau_heating_2020}, yet barely changes power sector costs. A complete coal phase-out in the electricity sector also does not have major effects on the impacts of accelerated heat pump rollouts on power sector capacities, dispatch, or costs. Higher natural gas prices have more substantial effects and, in particular, lead to higher additional power sector costs of additional heat pumps. Considering a week-long, pan-European renewable energy drought requires overall more firm capacities, and the rollout of heat pumps is accompanied by substantially higher solar PV investments in this case.

%Overall effects on natural gas and costs and emissions
We further find that an accelerated replacement of gas boilers with heat pumps can bring yearly natural gas savings between around 71 and 211~TWh\textsubscript{th}, depending on the rollout speed and gas prices, already accounting for increased gas usage in the electricity sector. For instance, in a \textit{fast} rollout to 10~million units in 2030, the additional heat pumps could save more than half of the private and commercial natural gas demand in Germany, which corroborates related findings \cite{altermatt_replacing_2023}. Overall yearly system cost savings depend, among other factors, on the natural gas price and range from around 2~to~27~billion~Euro for different natural gas price assumptions. \ch{CO2} emissions decrease by around 14-42 million~tons~per~year, corresponding to around 18-53\% of German households' carbon emissions from buildings in 2022.

% Discussion of limitations (and their qualitative effects on results?) and avenues for future work
As with any model-based analysis, our study has limitations. For example, we implicitly assume perfect distribution and transmission grids within countries that neglect any kind of grid congestion caused by heat pumps. In some distribution grid settings, the effect of heat pumps on grid congestion may be more severe than the impacts on system-wide generation capacities and dispatch modeled here. We also note that the hourly heat demand profiles used in our study are smoother than empirically measured heat pump operation patterns from the U.K. \cite{watson2021, ruhnau2023, halloran2024}. A ``peakier'' future heat demand pattern could potentially lead to higher load peaks and hence impact flexibility potentials of heat pumps, which merits further investigation in future work. In addition, our heat demand time series follow a synthetic test reference year approach, while the renewable electricity generation profiles and ambient temperatures come from actual weather years. This may lead to underestimating the system challenges in case of situations where very low renewable availability coincides with very low ambient temperatures and, accordingly, high heat demand. Therefore, it appears advisable to use consistent electric and heat load data as well as renewable generation profiles from the same weather years for future work. In addition, our approach of exogenously fixing the bioenergy capacity may lead to an underestimation of its flexibility potential. Without increasing the overall use of bioenergy, its conversion into electricity could become more concentrated in fewer hours to better complement variable wind and solar power. This would require a higher installed generation capacity (with lower full-load hours), as well as appropriate storage of biomass or biogas. 

Furthermore, Germany is not the only country pushing for an accelerated rollout of heat pumps. While we assume inflexible heat pumps outside Germany, future work could analyze rollouts in the whole of Europe in more detail to obtain more comprehensive insights into a wider European heating transition. Finally, our assumption of balanced charging for electric vehicles may not reflect reality. As the electric vehicle market evolves and charging infrastructure develops, there may be substantial changes in charging behavior and the adoption of smart and bidirectional charging technologies, which may decrease the value of flexibility provided by heat storage. Future research should explore more detailed modeling approaches that account for such potential changes. Likewise, smartly charged electric trucks could also be considered in future work \cite{gaetemorales2024}.

Our results show that even relatively small heat storage capacities may already have substantially positive power system effects. While in this analysis, heat pumps were either operated totally inflexibly with no heat storage or perfectly system-oriented with heat storage, future research could analyze the effects of other, and potentially more realistic, operating behaviors. The benefits of short-duration heat storage should be examined further with more volatile heat demand profiles to gain a more comprehensive understanding of its flexibility potential in energy systems. 
Further, operating heat pumps in a flexible way requires the right incentives for consumers. Hence, from a policy perspective, it is important to make sure that electricity consumption can be measured and controlled on a continuous basis (sometimes referred to as ``smart metering'') and that consumers have the possibility to choose electricity tariffs that reflect the dynamics of wholesale electricity markets. While very large heat storage sizes do not appear to be realistic for decentralized heat pumps, our results still serve as an indication of how larger, centralized heating systems with long-duration heat storage could operate.
As our analysis focuses primarily on the power sector effects of heat storage, the capacity of which is varied exogenously, future research might aim to investigate the optimal size of heat storage and its main influence factors.

% Final mic-drop
In summary, we find the power sector impacts of an accelerated heat pump rollout in Germany to be moderate and manageable, even under the assumption that the electric load from heat pumps has to be met by a corresponding yearly increase in renewable electricity generation. If wind energy expansion is restricted, additional solar PV capacity can be deployed instead without substantially increasing the overall system costs, facilitated by the European interconnection. In general, operating heat pumps in a temporally flexible manner entails substantial power sector benefits. Even relatively small heat storage already helps to reduce the additional needs for firm capacities or electricity storage induced by heat pumps and lowers power sector costs. To sum up, operating heat pumps in a temporally flexible manner is not strictly a ``must-have'' in the power sector modeled here, but it emerges as a desirable feature of the energy transition.

% ------------------------------------------------------------------
\section{Methods}
% ------------------------------------------------------------------

In the following, we describe the methodological approach as well as the sectoral and geographical scope of the study. First, we introduce the power sector model used in this analysis. Second, we provide details about the modeling of the heating sector and, in particular, the assumptions regarding the operation of heat pumps in our model. Third, we outline how other sector coupling options are considered in the model, namely electric mobility and the production of green hydrogen. Finally, we describe the geographical scope of the model. 

\paragraph{Power sector model} In this study, we use the power sector model DIETER (Dispatch and Investment Evaluation Tool with Endogenous Renewables), which has already been used in prior studies of energy storage and sector coupling \cite{schill2018results,schill_flexible_2020,roth_geographical_2023,kirchem2023hydrogen,gaetemorales2024}.\footnote{The model code can be accessed here: \url{https://gitlab.com/diw-evu/projects/heatpumps_2030}.} It is an open-source linear program that determines the least-cost investment and dispatch decisions for a range of electricity generation and storage technologies. The model minimizes total system costs while considering all subsequent hours of a year to capture renewable energy variability and storage use accurately. A detailed description of the objective function and the most relevant constraints can be found in \cite{zerrahn2017review}. The model covers the electricity sector and includes a detailed space heating module, e-mobility, and flexible hydrogen production options. Input data include time series of electric load, heat demand, electric vehicle charging, hydrogen demand, and capacity factors of renewable energies. Cost assumptions and technology investment constraints are further inputs. We are following a brown-field approach, in which we consider exogenous bounds on investments to account for existing plants and path dependencies. These are aligned with the current renewable capacity expansion plans of the German government and the currently installed fossil-fuel capacities. More detail on the capacity bounds can be found in section~\ref{subsec:data:electricity} and Table~\ref{tab:capacity_bounds}.

\paragraph{Heating sector} The German space heating sector is characterized by twelve archetypes of residential buildings categorized by two size classes (single-/two-family homes and multifamily buildings) and six age classes, corresponding to varying energy efficiency levels. While the building stock is described in detail in \cite{schill_flexible_2020}, we provide a brief overview here. We model twelve different building archetypes, which are distinguished by year of construction (six classes: before 1957, four periods between 1958 and 2019, and after 2019) and housing type (two classes: one -\& two-family homes and multifamily homes). Depending on the year of construction, the building archetypes are characterized by different energy efficiency levels: younger buildings have a lower annual heating requirement, and buildings constructed after 2020 are characterized as passive houses. Table~\ref{tab:building_archetypes} depicts the building stock assumptions for 2030, which are based on \cite{schill_flexible_2020}. 

For each of the twelve building archetypes, an hourly heating demand time series is generated using the open-source thermal building model TEASER (Tool for Energy Analysis and Simulation for Efficient Retrofit, \cite{remmen2018}) and the publicly available AixLib Library \cite{mueller2016}. The thermal building model considers the physical components of all major building elements and their thermal inertia to derive hourly heat flows inside the building and toward the ambiance (conduction, convection, and radiation). We assume indoor temperature requirements of 22$^{\circ}$C at daytime and a nighttime reduction to 18$^{\circ}$C between 10~p.m. and 5~a.m. Further, a test reference year approach is used to derive heating demand, which is representative of historical weather data in central Eastern Germany. Domestic hot water demand is modeled separately based on the Swiss SIA 2024 standard \cite{sia2006}. A graphic representation of the resulting hourly heat demand time series is provided in the Supplemental Information (Figure \ref{fig:heatmap heatdemand}). 

We exogenously set the share of total space heating and hot water demand, which has to be covered by two different types of heat pumps for each scenario; hence, we implicitly only consider the part of the building stock where heat pumps are installed. On an hourly basis, the heat pumps must satisfy the demand for space heating and hot water, as determined by the shares. We assume that heat pumps can be combined with buffer a heat storage of different sizes, which vary between scenarios. Based on these inputs and assumptions, the model optimizes the hourly electricity use by heat pumps.

\begin{figure}[H]
    \centering
    \caption{Heat module in DIETER}
    \includegraphics[width=0.65\textwidth]{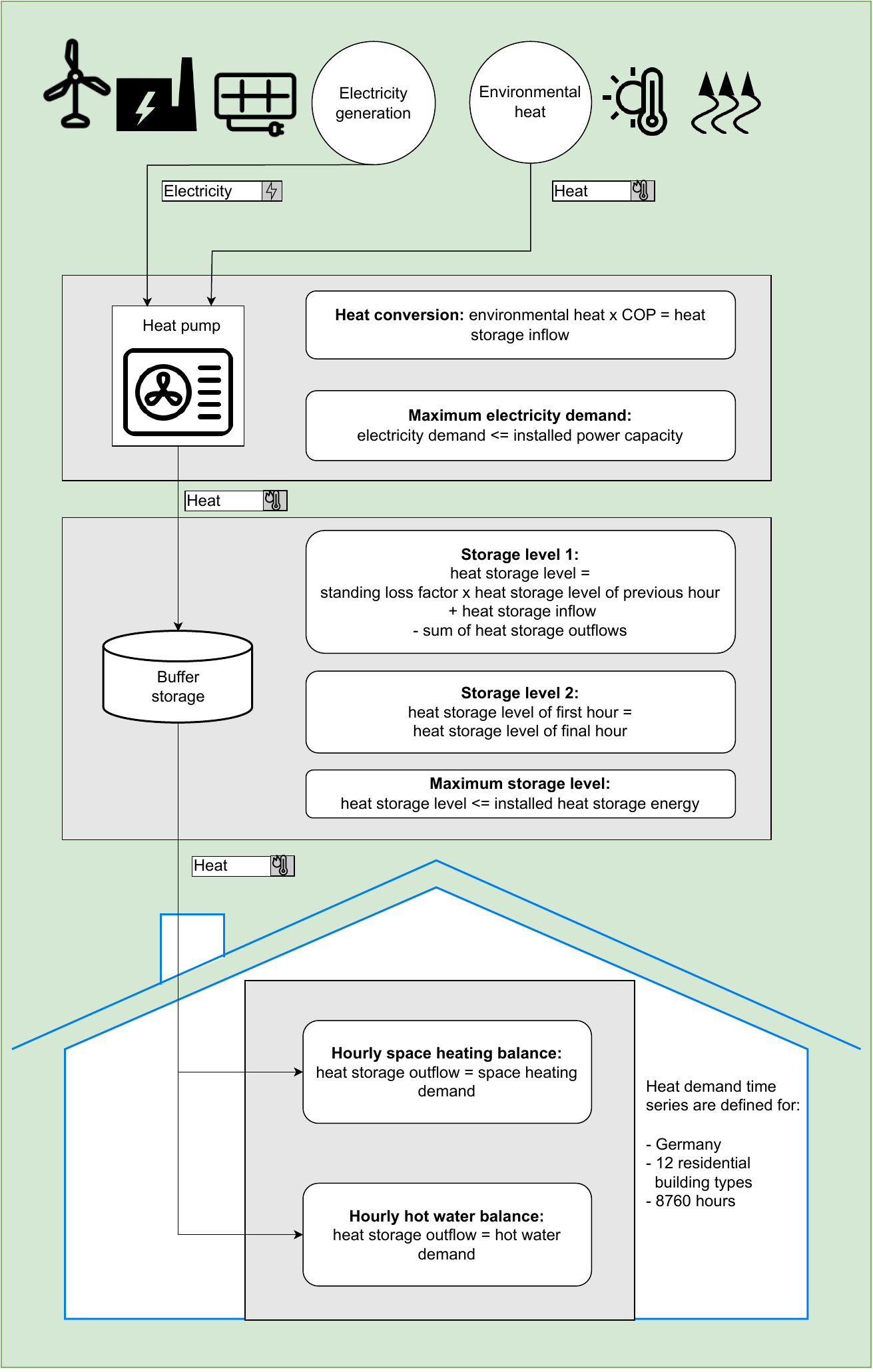}
    \label{fig:dieter_heatmodule}
\end{figure}

Figure \ref{fig:dieter_heatmodule} depicts how heat pumps are modeled in DIETER. The electric energy needed depends on the coefficient of performance (COP), which in turn depends on the ambient temperature in the case of air-source heat pumps. Lower ambient temperatures decrease the COP, so more electric energy is required to provide the same amount of heating energy. For more information, see section~\ref{SI:COP}. How much heating energy is provided to the building depends on the heat outflow from the buffer storage, which can neither exceed the total amount of heating energy stored plus the storage inflow in the same hour nor the installed heat output capacity of the heat pump. We only consider decentralized heat pumps with decentralized thermal energy storage. Centralized large heat pumps supplying district heating grids and centralized seasonal heat storage are not part of the analysis.

\paragraph{Other sector coupling options} As the electrification of other energy sectors is a policy target in Germany, we also account for electric mobility and the production of green hydrogen. The additional system load of electric vehicles enters the model as an electricity demand time series. Cars are assumed to charge with a balanced, yet not wholesale market price-driven time profile determined by the open-source tool ``emobpy'' \cite{gaete-morales_open_2021} (for further details, see~\ref{SI:bev}). The model also has to satisfy a given yearly demand for green hydrogen that has to be produced with electrolysis. The hourly hydrogen production profile is endogenously optimized, with given electrolysis capacity and assuming hydrogen storage at no cost. We provide the equations that describe the straightforward hydrogen model in Section~\ref{SI:hydrogen} in the Supplemental Information. 

\paragraph{Geographical scope} The study focuses on Germany, where an explicit heat pump rollout is modeled, but also includes Denmark, Poland, Czechia, Austria, Switzerland, France, Luxembourg, Belgium, the Netherlands, and Italy. To keep the model tractable while still considering the effects of European interconnection, we optimize investment decisions only in the German power sector while assuming (largely) fixed power plant fleets for other countries. We also do not explicitly model sector coupling for countries besides Germany. Section \ref{subsec:data:electricity} discusses capacity bounds for different countries.

% ------------------------------------------------------------------
%\section{Data and scenario assumptions}\label{sec:data and scenarios}
% ------------------------------------------------------------------

\subsection{Input data sources}

Time series data for the electric load, capacity factors for renewables, and hydro inflow data for all countries are taken from the ENTSO-E Pan-European Climate Database (PECD 2021.3) \cite{de_felice_2022}.\footnote{We use the target year 2030 and the weather year 2008.} Cost and technology parameters of electricity generation and storage technologies are depicted in Table~\ref{tab:costs} in the Supplemental Information. The relevant technical assumptions related to heating technologies and gas-based electricity generation technologies for the ex-post analysis of natural gas and emission saving are shown in Table~\ref{tab:gas_parameters} (more information in Section \ref{sec:gas and emission savings}).

\subsection{Scenario assumptions}

We refer to our main set of scenario assumptions as \textit{baseline}. In the following, we briefly sketch the most important features of this scenario. Whenever we deviate from the baseline, for instance, when we present sensitivity analyses, we make this explicit.

\paragraph{Heating sector}

We distinguish between three policy scenarios of the overall heat pump stock in 2030 which we compare to a \textit{reference} scenario. In this \textit{reference} scenario, we assume 1.7~million decentralized heat pumps in 2030. This number reflects the stock of heat pumps installed in Germany at the time of writing, meaning no further pumps would be installed until 2030. In the \textit{slow} rollout, the number of heat pumps would reach 3~million by 2030. Additional heat pumps would be exclusively installed in single- and two-family homes built between 1995 and 2009. This scenario largely corresponds to the current growth of heat pump deployment in Germany.\footnote{See \url{https://openenergytracker.org/en/docs/germany/heat/}}
In the \textit{government} rollout, the stock of heat pumps would reach 6~million in 2030, reflecting the target of the German government \cite{schill_mixed_2023}. In this scenario, most single- and two-family homes built after 1995 would be equipped with heat pumps. In the \textit{fast} rollout, heat pumps would be installed in more single- and two-family homes, even old ones built before 1979 with very low energy efficiency standards. In this scenario, the total number of heat pumps would increase to 10~million by 2030, and the total annual heat provided by heat pumps increases substantially. Table \ref{tab:hp_stocks} provides an overview of the different heat pump rollout scenarios. Table~\ref{tab:building_archetypes} provides additional information on how heat pumps are rolled out across different building types. In the most ambitious scenario, decentralized heat pumps provide around 40\% of total space heating and domestic hot water needs (Table~\ref{tab:hp_stocks}). The electricity demand of heat pumps and other sectors in the different scenarios is depicted in Figure~\ref{fig:electricity demand}.

Across all building types, air-source heat pumps account for 80\% of installed heat pumps and ground-source heat pumps account for the remaining 20\%. While ground-source heat pumps are more energy-efficient, air-source heat pumps are cheaper to install. We assume that all heat pumps can, in principle, be combined with thermal energy storage. We conduct analyses with varying storage energy capacities. The assumed energy storage size is expressed in energy-to-power (E/P) ratios ranging from zero to 168 hours (0,~2,~6,~24, and 168~hours). In this terminology, a heat storage with an E/P ratio of 2 hours has a total heat storage capacity that equals 2 hours of maximum heat output of the heat pump. Equipping heat pumps with a 0-hour storage means that heat pumps have no attached heat storage and thus have to exactly follow the heat demand profile in every hour. That means, heat pumps are operated inflexibly. With increasing heat storage, heat pumps can be operated with more flexibility, allowing to decouple electricity consumption from heat provision. Importantly, our modeling approach assumes that heat pumps are operated in a system-friendly, i.e.,~cost-minimizing manner whenever possible. This could be interpreted as if heat pump operators faced hourly wholesale prices and operated their heat pumps to minimize overall system costs. As this is not the case today, we discuss the consequences of this assumption in the conclusion. 

\paragraph{Capacity bounds}\label{subsec:data:electricity}

In Germany, we limit the capacities of coal- and oil-fired power plants to current levels. Capacities of gas-fired power plants, open-cycle~(OCGT), and combined-cycle~(CCGT) - following current policy discussions - can be expanded beyond current levels. In sensitivity analyses with a German coal phase-out, we assume the upper capacity limit for hard coal and lignite to be zero. Regarding wind energy, we align upper capacity bounds for on- and offshore wind energy with the current German government targets of 115~GW for onshore wind and 30~GW for offshore wind in the baseline scenarios. We use the government target for wind power as an upper limit, as wind capacity expansion in Germany is slow due to long assessment and permit processes, and limited by land (and sea) availability. In this context, the government targets can be considered ambitious. An even higher wind energy capacity expansion appears unrealistic to achieve by 2030~\cite{schill_mixed_2023}. In a sensitivity analysis, we remove these upper bounds on wind power capacities. The capacities of solar PV do not have any bounds. The electrolysis capacity is fixed at 10~GW\textsubscript{e}. 

In other countries, renewable energy capacities are fixed based on the values of Ten-Year Network Development Plan (TYNDP)~\cite{entsoe_tyndp_2018} of ENTSO-E and set as upper bounds for fossil generators. The reason for not fixing the fossil generators in other countries is to avoid an unduly large power plant fleet that could support the German heat pump rollout. Therefore, the model is free to choose the smallest capacity needed. In all countries, we fix the capacities of all hydropower technologies (run-of-river, reservoirs, and pumped-hydro) according to the ERAA~2021 \cite{de_felice_2022} and bioenergy under the assumption that their potential for further capacity expansion is exhausted. Table~\ref{tab:capacity_bounds} provides an overview of all capacity bounds in all countries.

\paragraph{Sector coupling demand}

While we assume an annual conventional load of 550~TWh in Germany in all scenarios, we additionally consider new electric loads related to electric mobility and hydrogen production. To incorporate the impact of electric mobility, we include a fleet of 15~million electric cars compatible with the government's goal of 2030 \cite{schill_mixed_2023}. This fleet would require approximately 36~TWh of additional electricity annually. Figure \ref{fig:ev_timeseries} depicts hourly demand patterns of that fleet. Additionally, we account for 28~\ch{TWh_{H_2}} of hydrogen demand in Germany produced by domestic electrolysis, resulting in an additional electricity demand of around 39~TWh. This number is based on the target set in the updated German National Hydrogen Strategy of 2023 to build up an electrolysis capacity of 10~GW \cite{nws2023, kittel2023national} and assuming 4000~full-load hours. This does not include hydrogen imports, which are expected to satisfy 50~to~70~\% of the German hydrogen demand (95-130 \ch{TWh_{H_2}} in 2030) \cite{nws2023}. We further assume that hydrogen can be stored without additional investment costs, e.g.,~in existing cavern storage. This enables electrolyzers to operate with a substantial degree of flexibility to produce hydrogen over the course of the year. In countries besides Germany, additional loads related to sector coupling are included in the electric load time series data provided by ENTSO-E and thus assumed to be inflexible. Figure~\ref{fig:electricity demand} provides an overview of the electricity demand of the different sector coupling options.

\paragraph{Renewable energy constraint}

In all scenarios, 80\% of the yearly electricity consumption in Germany (including the consumption of electric vehicles and electrolysis) has to be covered by renewable energy sources. That is in line with the goal of the current German government coalition. In addition, the electricity demand by heat pumps has to be entirely met by additional renewable energy sources over the course of a year (but not in every single hour). In other countries, we do not assume any renewable energy targets.

\paragraph{Fuel and carbon prices}

Our fuel price assumptions are summarized in Table \ref{tab:costs}. In our \textit{baseline} assumptions, we set the wholesale price of natural gas to 50~Euro per MWh. We further assume a carbon emission cost of 130~Euro~per~ton~of~CO\textsubscript{2} in 2030 \cite{pietzcker_notwendige_2021}. This cost is associated with the emission factor of fossil-based heating and electricity generation technologies and is considered a variable generating cost, along with fuel expenses.

% -------------------------
    
\section*{Acknowledgments}

We thank Oliver Ruhnau and two anonymous reviewers for very helpful comments, as well as various colleagues of the Ariadne project for feedback on earlier drafts. We thank our colleague Adeline Gu\'{e}ret for supporting the calculations described in section~\ref{sec:gas and emission savings}. We further thank the participants of the \textit{IAEE 2023 Conference Milano}, the \textit{Smart Energy Systems - International conference 2023 Copenhagen}, and the \textit{Strommarkttreffen 2024/01 Cologne} for their constructive feedback. 

We gratefully acknowledge financial support from the German Federal Ministry of Education and Research (BMBF) via the Kopernikus project Ariadne (FKZ 03SFK5N0, FKZ 03SFK5N0-2), as well as from the Federal Ministry of Labour and Social Affairs (BMAS) via the project  FIS (FIS.03.00016.21).

% ------------------------------

\section*{Author contributions}

\textbf{Conceptualization:} AR, WS;
\textbf{Methodology:} AR, CG, DK, WS;
\textbf{Software:} AR, CG;
\textbf{Formal analysis:} AR, DK, WS;
\textbf{Investigation:} AR;
\textbf{Data Curation:} AR, CG, DK, WS;
\textbf{Writing - Original Draft:} AR, DK;
\textbf{Writing - Review \& Editing:} WS;
\textbf{Visualization:} AR;
\textbf{Supervision:} WS;
\textbf{Funding acquisition:} WS

% ----------------------

\section*{Data availability}

The input and results data used in this paper can be accessed here: \url{https://zenodo.org/records/13844622}.

% ----------------------

\section*{Code availability}

The model and analysis code used in this paper can be accessed here: \url{https://zenodo.org/records/13844622}.

% ----------------------

\section*{Ethics declarations}

\section*{Competing interests}

The authors declare no competing interests.

% ----------------------

\newpage

% ----------------------

\newpage

\onehalfspacing

%% Bibliography
%\bibliographystyle{unsrt} 
%\bibliography{main}
\printbibliography

% ----------------------

\newpage

% Figures

% Tables ---------------------------
\newpage

\begin{table}[H]
    \centering
    \caption{Heat pump data}
    \begin{tabular}{llrrrr}
    \hline
    &   & \textbf{Reference} & \textbf{Slow} & \textbf{Government} & \textbf{Fast} \\
    \hline
    Number of installed heat pumps     & [million]               & 1.7   & 3.0   & 6.0   & 10.0  \\
    Heat pump power rating             & [GW\textsubscript{e}]   & 8.7   & 14.5  & 27.5  & 52.6  \\
    Maximum thermal heat pump output           & [GW\textsubscript{th}]  & 19.6  & 32.7  & 61.9  & 118.5 \\
    Share of air-sourced heat pumps    &                         & 0.8   & 0.8   & 0.8   & 0.8   \\
    Share of ground-sourced heat pumps &                         & 0.2   & 0.2   & 0.2   & 0.2   \\
    Yearly heat supplied by heat pumps & [TWh\textsubscript{th}] & 24.7 & 53.2  & 92.9 & 226.3 \\
    \hline
    \end{tabular}
    \begin{minipage}[c]{\textwidth}
    \bigskip
    \footnotesize
    Heat includes space heating and domestic hot water.
    \end{minipage}%
  \label{tab:hp_stocks}%
\end{table}%

% Table generated by Excel2LaTeX from sheet 'summary results'
\begin{table}[H]
\centering
\caption{Yearly saving of natural gas, \ch{CO2_{eq}} emissions, and costs related to heat pumps}
\resizebox{\textwidth}{!}{
\begin{tabular}{p{12em}lcccccc}
\toprule
Gas price          & Euro/MWh  & \multicolumn{2}{c}{50}                              & \multicolumn{2}{c}{100}                             & \multicolumn{2}{c}{150} \\
Heat pump rollout &           & \multicolumn{1}{c}{Gov.} & \multicolumn{1}{c}{Fast} & \multicolumn{1}{c}{Gov.} & \multicolumn{1}{c}{Fast} & \multicolumn{1}{c}{Gov.} & \multicolumn{1}{c}{Fast} \\
\midrule
%    Gas displaced for heating             & TWh\textsubscript{th}        & -51.0 & -158.4 & -51.0 & -158.4 & -51.0 & -158.4 \\
Natural gas displaced by additional heat pumps   & TWh\textsubscript{th}                 & -75.75	       & -223.97	      & -75.75	        & -223.97	      & -75.75	        & -223.97 \\
Additional gas usage for electricity  generation & TWh\textsubscript{th}                 & +2.58	       & +18.01	          & +1.92	        & +12.75	      & +4.90	        & +20.35\\
Total gas savings                       & TWh\textsubscript{th}        & -73.17 & -205.96 & -73.83 & -211.22 & -70.85 & -203.62 \\
Total emissions savings                 & Mio t \ch{CO2_{eq}}          & -14.63 & -41.19  & -14.77  & -42.24 & -14.17  & -40.72  \\
Change in overall system costs          & Billion EUR                  & -2.05  & -6.73   & -5.48 & -16.80  & -9.07  & -27.07  \\
\bottomrule
\end{tabular}%
}
\par
\begin{minipage}[c]{\textwidth}
\bigskip
\footnotesize
    Changes and savings are shown relative to the respective reference scenario. \textit{Gov.} refers to the \textit{government} rollout scenario of heat pumps.
\end{minipage}
\label{tab:gas_savings}%
\end{table}%

% ----------------------

\newpage

\appendix

\renewcommand{\thesection}{SI}
\renewcommand{\thepage}{SI}
\global\long\def\thefigure{SI.\arabic{figure}}
\global\long\def\thetable{SI.\arabic{table}}
\global\long\def\thepage{SI.\arabic{page}}
\setcounter{figure}{0}
\setcounter{table}{0}
\setcounter{page}{1}

\newcommand{\invisiblesection}[1]{%
  \phantomsection%
  \stepcounter{section}%
  \addcontentsline{toc}{section}{\protect\numberline{\thesection}#1}%
  }

\newpage

\thispagestyle{plain}

\invisiblesection{Supplemental Information}

\begin{center}
    \par \vspace{2cm}
    {\Huge Power sector benefits of flexible heat pumps} \par
    \vspace{2cm}
    {\Large \textbf{Supplementary Information}} \par
    \vspace{2cm}
    {\large Alexander Roth, Carlos Gaete-Morales, Dana Kirchem, Wolf-Peter Schill}
\end{center}

\newpage

\subsection{Supplementary Methods}

This section provides additional information regarding the model we use for our analysis.

\subsubsection{Electricity demand}

\begin{figure}[H]
    \centering
        \caption{Electricity demand Germany}
        \includegraphics[width=0.99\textwidth]{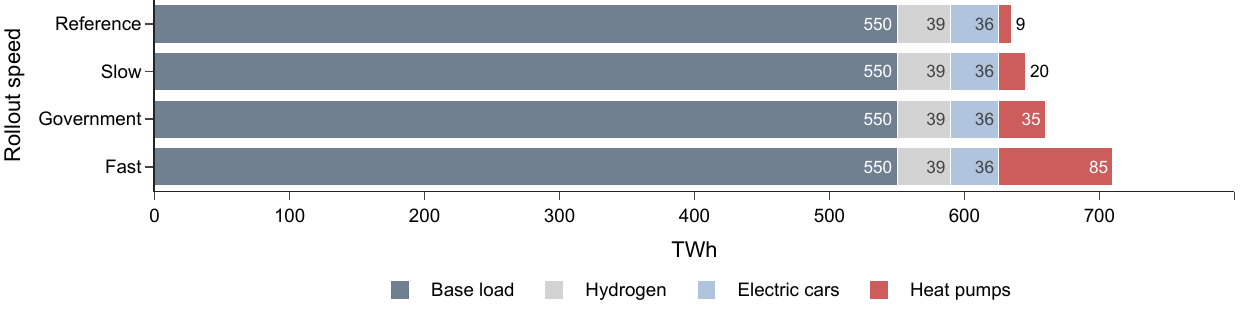}
        \begin{minipage}[c]{\textwidth}
        \bigskip
        \footnotesize
        Electricity demand for those scenarios are shown in which heat pumps do not have heat storage and hence are operated inflexibly.
        \end{minipage}
        \label{fig:electricity demand}
\end{figure}

Figure \ref{fig:electricity demand} provides an overview of the different sources of electricity demand in our model runs.

\subsubsection{Heat pumps} \label{SI:COP}

Following previous research \cite{schill_flexible_2020}, we model the coefficient of performance (COP) of heat pumps as follows:

\begin{equation}
\text{COP}^{hp}_{h} = \eta^{\text{hp}} \frac{{\text{temp}^{\text{sink}} + 273.15^\circ \text{C}}}{{\text{temp}^{\text{sink}} - \text{temp}_{h}^{\text{source}}}}
\end{equation} 

We generally assume a sink temperature of 50$^\circ$C. For ground-source heat pumps, we assume a constant source temperature $\text{temp}_{h}^{\text{source}}$ of 10$^\circ$C for all hours. Assuming a dynamic efficiency parameter $\eta^{\text{ground}}$ of $0.45$, this renders a COP of 3.64 for ground-sourced heat pumps. For air-source heat pumps, the source temperature varies with the hourly ambient temperature. Using an efficiency parameter $\eta^{\text{air}}$ of $0.35$, this renders a COP of 2.83 in hours with an ambient temperature of 10$^\circ$C, and a COP of 2.26 at 0$^\circ$C. Compared to state-of-the-art heat pumps, these values may be considered conservative, resulting in inflated electricity consumption of heat pumps. Yet, we assume that heat pumps are also installed in older buildings where heating systems may require higher sink temperatures, which would decrease the COP, especially in the fast rollout scenario. As we assume a constant sink temperature of 50$^\circ$C in all building types, this makes our COP assumptions appear less conservative.

\subsubsection{Electric vehicles} \label{SI:bev}

\begin{figure}[H]
\centering
    \caption{Hourly average electricity demand of 15 million BEV for a exemplary week.}
    \includegraphics[width=0.9\textwidth]{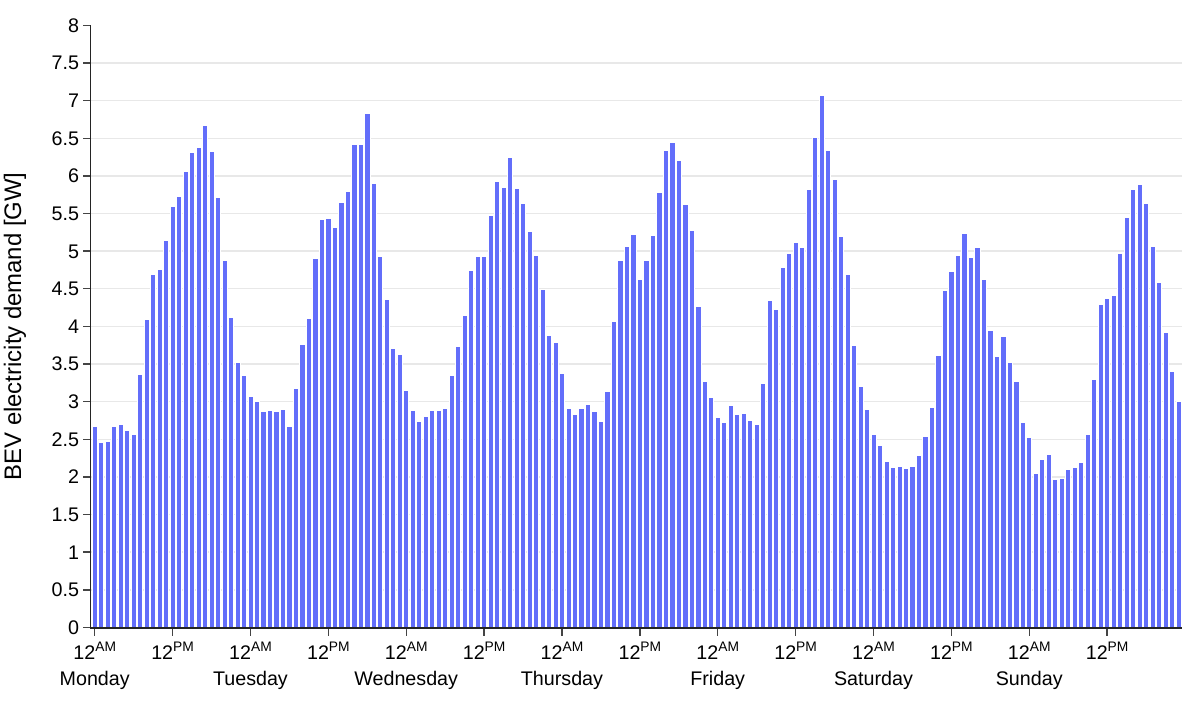}
    \label{fig:ev_timeseries}
\end{figure}

The analysis includes battery electric vehicle (BEV) time series using the emobpy tool \cite{gaete-morales_open_2021}. The dataset \cite{gaete-morales_emobpy_2021} used has been created utilizing data from the ``Mobilität in Deutschland'' survey, distinguishing between commuter and spontaneous drivers and incorporating various factors such as trip frequencies, distances, trip duration, departure times, charging station availability, and charging strategies, as well as the use of popular BEV models. The dataset encompasses multiple charging strategies. For this research, we have selected the ``immediate-balanced'' approach to reflect the electricity drawn from the grid. Under this charging strategy, the vehicles' batteries are charged upon arriving at charging stations, with a constant and often lower power rating than the charging station. This approach ensured that the BEV reached a 100\% state of charge just before commencing the next trip. The selected time series are scaled to represent the demand for 15~million battery electric vehicles, with an annual electricity demand of 36~TWh. Figure \ref{fig:ev_timeseries} depicts the electricity demand of BEV in an exemplary week. For more information regarding the construction of the BEV demand time series, we refer to previous work \cite{gaete-morales_emobpy_2021,gaete-morales_open_2021}.

\subsubsection{Green hydrogen} \label{SI:hydrogen}

In the analysis, the production of green hydrogen is only foreseen in Germany and modeled following the approach described in \cite{zerrahn_2018}. We assume that a given hydrogen demand $h2^{demand}$ of 28~TWh has to be covered by electrolysis over the course of a year (Equation~\ref{eq:h2_model1}). We assume a temporally flexible hydrogen demand and unlimited hydrogen storage. The electrolysis capacity is exogenously set to 10~GW in Germany (Equation~\ref{eq:h2_model3}). The conversion factor of electrolysis is 71\%; hence, one~kilowatt-hour~(kWh) of electricity is transformed into 0.71~kWh of hydrogen.

\begin{align}
    \label{eq:h2_model1} h2^{demand} = \sum^h H2^{prod}_h \\%sum(h, h2_time_data(h,n)) =E= sum(h, H2_H2_PRODUCTION(n,h))
    \label{eq:h2_model2} H2^{prod}_h = H2^{elec}_h \times 0.71 \\%H2_H2_PRODUCTION(n,h) =E= H2_ELEC_IN(n,h) * 0.71 * 1000
    \label{eq:h2_model3} H2^{elec}_h \leq INV^{H2} %H2_ELEC_IN(n,h)      =L= H2_ELECTROLYSIS(n)
\end{align}

\newpage

\subsubsection{Input data}

In this section, we present the input data to our model in more detail. 

\vspace{1cm}

\begin{table}[H]
  \centering
  \caption{Building archetypes and heating energy demand assumptions for Germany in 2030}
  \resizebox{\textwidth}{!}{
    \begin{tabular}{ll|ccc|cccccccc}
    \hline
    && \textbf{Total number} & \textbf{Annual} && \multicolumn{8}{c}{\textbf{Shares of heat pumps [\%]}} \\
    \textbf{Type} & \textbf{Year construction} & \textbf{of buildings} & \textbf{heating energy}             & \textbf{Floor area} & \multicolumn{2}{c}{\textbf{Reference}} & \multicolumn{2}{c}{\textbf{Slow}} & \multicolumn{2}{c}{\textbf{Government}} & \multicolumn{2}{c}{\textbf{Fast}}\\
                  &                            & [million]                       & \textbf{demand} [kWh/m\textsuperscript{2}] & [million m\textsuperscript{2}] & air & ground & air & ground & air & ground & air & ground \\
    \hline
      & \textbf{One- \& two-family houses}  &       &       &     & & & & & & & & \\
    1 & \hspace{0.2cm} Before 1957          & 1.41  & 276   & 247 & 0.008  & 0.002  & 0.008   & 0.002   & 0.008   & 0.002   & 0.2768 & 0.0692 \\
    3 & \hspace{0.2cm} 1958-1978            & 2.46  & 203   & 431 & 0.008  & 0.002  & 0.008   & 0.002   & 0.008   & 0.002   & 0.2768 & 0.0692 \\
    5 & \hspace{0.2cm} 1979-1994            & 2.55  & 153   & 446 & 0.0136 & 0.0034 & 0.0136  & 0.0034  & 0.0136  & 0.0034  & 0.72   & 0.18 \\
    6 & \hspace{0.2cm} 1995-2009            & 3.02  & 112   & 528 & 0.0488 & 0.0122 & 0.38288 & 0.09572 & 0.60048 & 0.15012 & 0.72   & 0.18 \\
    7 & \hspace{0.2cm} 2010-2019            & 1.75  & 66    & 306 & 0.272  & 0.068  & 0.272   & 0.068   & 0.72    & 0.18    & 0.72   & 0.18 \\
    9 & \hspace{0.2cm} After 2019           & 2.15  & 15    & 375 & 0.272  & 0.068  & 0.272   & 0.068   & 0.72    & 0.18    & 0.72   & 0.18 \\
      & \textbf{Multifamily houses }        &       &       &     & & & & & & & &\\ 
    2 & \hspace{0.2cm} Before 1957          & 0.34  & 223   & 170 & 0.0104 & 0.0026 & 0.0104 & 0.0026 & 0.0104 & 0.0026 & 0.0104 & 0.0026 \\
    4 & \hspace{0.2cm} 1958-1978            & 0.64  & 164   & 322 & 0.0104 & 0.0026 & 0.0104 & 0.0026 & 0.0104 & 0.0026 & 0.0104 & 0.0026 \\
    6 & \hspace{0.2cm} 1979-1994            & 0.46  & 130   & 230 & 0      & 0      & 0      & 0      & 0      & 0      & 0      & 0      \\
    8 & \hspace{0.2cm} 1995-2009            & 0.47  & 103   & 239 & 0.0112 & 0.0028 & 0.0112 & 0.0028 & 0.0112 & 0.0028 & 0.0112 & 0.0028 \\
    10& \hspace{0.2cm} 2010-2019            & 0.36  & 51    & 181 & 0.128  & 0.032  & 0.128  & 0.032  & 0.128  & 0.032  & 0.128  & 0.032  \\
    12& \hspace{0.2cm} After 2019           & 0.46  & 11    & 232 & 0.128  & 0.032  & 0.128  & 0.032  & 0.128  & 0.032  & 0.128  & 0.032  \\
    \hline
    \end{tabular}%
  }
  \label{tab:building_archetypes}%
\end{table}%

\noindent Table~\ref{tab:building_archetypes} provides information on the different building archetypes and their respective heating energy demand.

\newpage

\begin{figure}[H]
    \centering
    \caption{Space heating demand}
    \includegraphics[width=\textwidth]{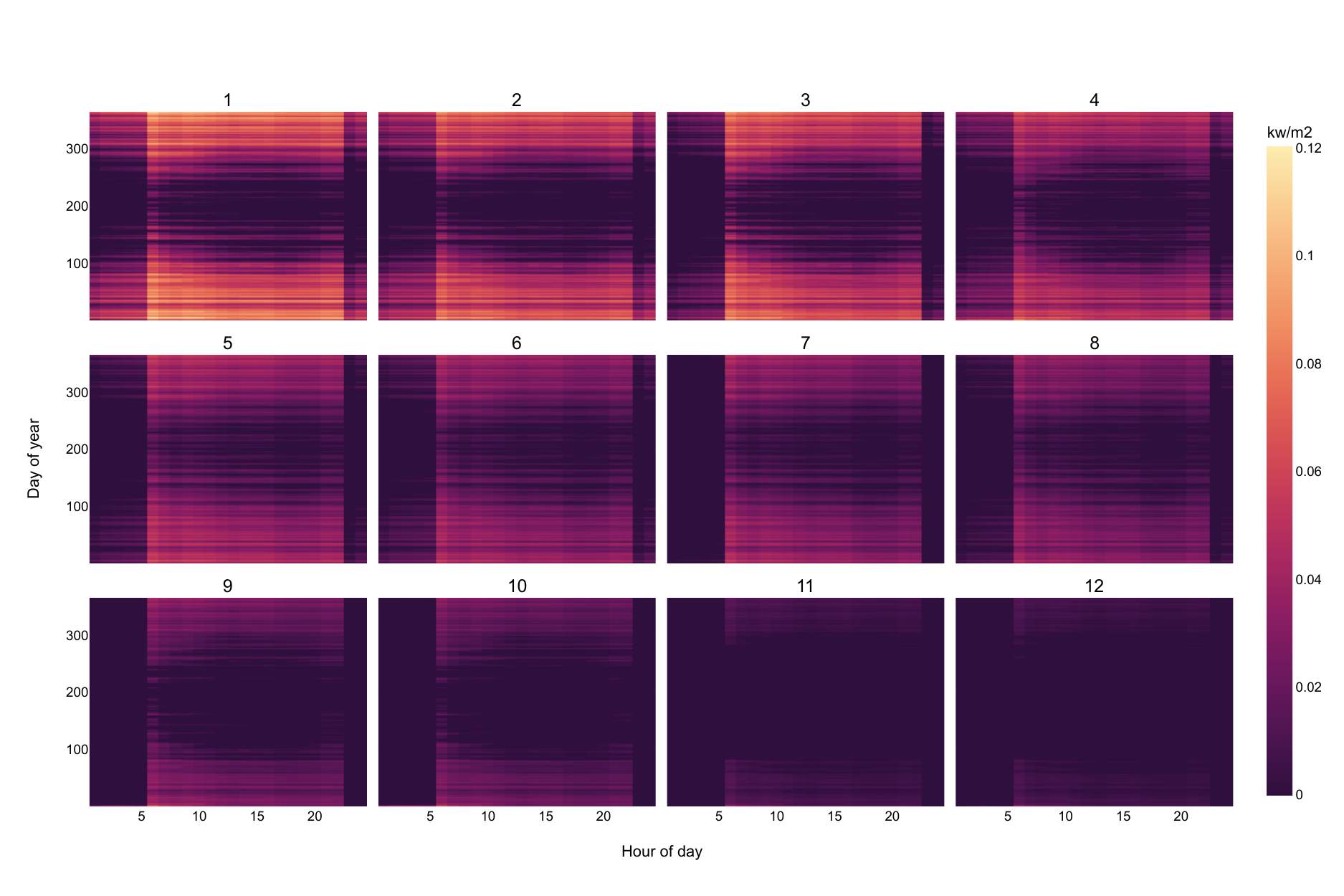}
    \begin{minipage}[c]{0.9\textwidth}
        \bigskip
        \footnotesize
        The number above every panel refers to the building archetypes, as defined in table \ref{tab:building_archetypes}.
    \end{minipage}
    \label{fig:heatmap heatdemand}
\end{figure}

\noindent Figure \ref{fig:heatmap heatdemand} depicts the heat demand in kWh/m$^2$ for different housing classes, depending on the hour of the day and the day of the year.

\newpage

\begin{table}[H]
\centering
\caption{Assumptions on capacity bounds [GW]}
\resizebox{\textwidth}{!}{
\begin{tabular}{lrrrrrrrrrrr}
    \toprule
                                & \textbf{Germany}                 & \textbf{Austria}    & \textbf{Belgium}    & \textbf{Switzerland} & \textbf{Czech Republic} & \textbf{Denmark}    & \textbf{France}     & \textbf{Italy}      & \textbf{Luxembourg} & \textbf{Netherlands} & \textbf{Poland}     \\
    \midrule
    Run-of-river hydro          & 3.93                    & 6.38       & 0.15       & 4.22        & 0.43           & 0.00       & 13.60      & 7.03       & 0.04       & 0.04        & 0.37       \\
    Nuclear                     & 0.00                    & 0.00       & 0.00       & 1.19        & 4.04           & 0.00       & 58.21      & 0.00       & 0.00       & 0.49        & 0.00       \\
    Lignite                     & 0.00-9.30 (0.00)        & 0.00       & 0.00       & 0.00        & 0.00-3.89      & 0.00       & 0.00       & 0.00       & 0.00       & 0.00        & 0.00-6.32  \\
    Hard coal                   & 0.00-9.80 (0.00)        & 0.00       & 0.00-0.62  & 0.00        & 0.00-0.37      & 0.00-0.77  & 0.00       & 0.00       & 0.00       & 0.00        & 0.00-9.88  \\
    Natural gas (CCGT)          & 0.00-$\infty$           & 0.00-2.82  & 0.00-7.61  & 0.00        & 0.00-1.35      & 0.00       & 0.00-6.55  & 0.00-38.67 & 0.00       & 0.00-8.65   & 0.00-5.00  \\
    Natural gas (OCGT)          & 0.00-$\infty$           & 0.00-0.59  & 0.00-1.08  & 0.00        & 0.00           & 0.00       & 0.00-0.88  & 0.00-5.40  & 0.00       & 0.00-0.64   & 0.00       \\
    Oil                         & 0.00-1.20               & 0.00-0.17  & 0.00       & 0.00        & 0.00-0.01      & 0.00       & 0.00       & 0.00       & 0.00       & 0.00        & 0.00       \\
    Other                       & 0.00-4.10               & 0.95       & 1.32       & 0.89        & 1.23           & 0.24       & 1.87       & 5.99       & 0.03       & 3.77        & 6.82       \\
    Bio energy                  & 6.00                    & 0.60       & 0.21       & 1.20        & 1.06           & 0.67       & 2.56       & 4.93       & 0.05       & 0.54        & 1.41       \\
    Onshore wind                & 56.00-115.00 ($\infty$) & 10.00      & 5.93       & 1.25        & 3.00           & 5.48       & 44.11      & 19.05      & 0.35       & 8.30        & 11.28      \\
    Offshore wind               & 7.77-30.00 ($\infty$)   & 0.00       & 4.30       & 0.00        & 0.00           & 4.78       & 3.00       & 0.60       & 0.00       & 6.72        & 0.90       \\
    Solar PV                    & 59.00-$\infty$          & 15.00      & 13.92      & 11.00       & 10.50          & 4.75       & 42.63      & 49.33      & 0.25       & 15.46       & 12.19      \\
    Lithium-ion batteries       &                         &            &            &             &                &            &            &            &            &             &            \\
    ... power in/out            & 0-$\infty$/0-$\infty$              & 0-$\infty$/0-$\infty$ & 0-$\infty$/0-$\infty$ & 0-$\infty$/0-$\infty$  & 0-$\infty$/0-$\infty$     & 0-$\infty$/0-$\infty$ & 0-$\infty$/0-$\infty$ & 0-$\infty$/0-$\infty$ & 0-$\infty$/0-$\infty$ & 0-$\infty$/0-$\infty$  & 0-$\infty$/0-$\infty$ \\
    ... energy [GWh]            &  0-$\infty$              &  0-$\infty$ &  0-$\infty$ &  0-$\infty$  &  0-$\infty$     &  0-$\infty$ &  0-$\infty$ &  0-$\infty$ &  0-$\infty$ & 0-$\infty$  &  0-$\infty$ \\
    Power-to-gas-to-power       &                         &            &            &             &                &            &            &            &            &             &            \\
    ... power in/out            & 0-$\infty$/0-$\infty$              & 0-$\infty$/0-$\infty$ & 0-$\infty$/0-$\infty$ & 0-$\infty$/0-$\infty$  & 0-$\infty$/0-$\infty$     & 0-$\infty$/0-$\infty$ & 0-$\infty$/0-$\infty$ & 0-$\infty$/0-$\infty$ & 0-$\infty$/0-$\infty$ & 0-$\infty$/0-$\infty$  & 0-$\infty$/0-$\infty$ \\
    ... energy [GWh]            &  0-$\infty$              &  0-$\infty$ &  0-$\infty$ &  0-$\infty$  &  0-$\infty$     &  0-$\infty$ &  0-$\infty$ & 0-$\infty$ &  0-$\infty$ &  0-$\infty$  &  0-$\infty$ \\
    Open pumped hydro storage   &                         &            &            &             &                &            &            &            &            &             &            \\
    ... power in/out            & 1.86/2.14               & 5.33/5.61  & 0.00/0.00  & 1.89/2.46   & 0.60/0.65      & 0.00/0.00  & 1.85/1.85  & 2.22/3.62  & 0.00/0.00  & 0.00/0.00   & 0.17/0.22  \\
    ... energy [GWh]            & 471.23                  & 1746.66    & 0.00       & 1194.00     & 2.95           & 0.00       & 90         & 289.99     & 0.00       & 0.00        & 1.31       \\
    Closed pumped hydro storage &                         &            &            &             &                &            &            &            &            &             &            \\
    ... power in/out            & 7.17/7.01               & 0.45/0.45  & 1.23/1.31  & 1.90/1.90   & 0.64/0.69      & 0.00/0.00  & 1.95/1.95  & 4.17/4.17  & 0.00/0.00  & 0.00/0.00   & 1.49/1.33  \\
    ... energy [GWh]            & 391.59                  & 3.60       & 5.80       & 56.00       & 3.70           & 0.00       & 10.00      & 61.20      & 0          & 0           & 6.35       \\
    Reservoirs                  &                         &            &            &             &                &            &            &            &            &             &            \\
    ... power out               & 0.82                    & 2.80       & 0.00       & 8.53        & 0.52           & 0.00       & 9.8        & 8.77       & 0.00       & 0.00        & 0.42       \\
    ... energy [TWh]            & 0.24                    & 0.77       & 0.00       & 7.91        & 0.01           & 0.00       & 10.00      & 5.57       & 0.00       & 0.00        & 0.001      \\
    Electrolysis                & 10.00                   & 0.00       & 0.00       & 0.00        & 0.00           & 0.00       & 0.00       & 0.00       & 0.00       & 0.00        & 0.00       \\
    \bottomrule
\end{tabular}%
}
\begin{minipage}[c]{0.97\textwidth}
    \bigskip
    \footnotesize
    Based on Bundesnetzagentur \cite{bundesnetzagentur_genehmigung_2018} and ENTSO-E \cite{entsoe_tyndp_2018}. If two numbers are connected with a hyphen, the model can choose endogenously in that range; otherwise, the value is fixed. All values refer to the \textit{baseline} assumption. In brackets, numbers are provided for sensitivity analyses. All numbers are provided in GW except for storage energy, which is provided in GWh or TWh.
\end{minipage}
\label{tab:capacity_bounds}%
\end{table}

\noindent Table \ref{tab:capacity_bounds} provides an overview of the capacity assumptions and bounds of the different technologies in all countries.

\newpage

\vspace{1cm}

\begin{table}[H]
  \centering
  \caption{Cost and technical parameters for savings calculation}
    \begin{tabular}{l r}
    \toprule
    \textbf{Parameter} & \textbf{Value} \\
    \midrule
    Overnight investment costs [EUR/kilowatt\textsubscript{th}]              &        \\
    \hspace{0.5cm} \textit{Air-sourced heat pumps}                     & 850    \\
    \hspace{0.5cm} \textit{Ground-sourced heat pumps}                  & 1400   \\
    \hspace{0.5cm} \textit{Gas boilers}                                & 296    \\
    Efficiencies                                                       &        \\
    \hspace{0.5cm} \textit{Open-cycle gas turbine}                     & 0.4    \\
    \hspace{0.5cm} \textit{Combined-cycle gas turbine}                 & 0.542  \\
    \hspace{0.5cm} \textit{Gas boilers}                                & 0.9    \\
    Technical lifetime of heat pumps [Years]                           & 20     \\
    Interest rate                                                      & 0.04   \\
    Emission factor [t \ch{CO2_{eq}} / MWh\textsubscript{th}]  & 0.2    \\
    \bottomrule
    \end{tabular}%
  \label{tab:gas_parameters}%
\end{table}%

\noindent Table \ref{tab:gas_parameters} provides an overview of the cost and technical parameters for the calculation of gas, emissions, and cost savings.

\newpage

\begin{table}[H]
\centering
\caption{Cost and technology parameters}
\begin{subtable}[t]{\textwidth}
\caption{Electricity storage}
\tiny

%\setlength{\tabcolsep}{2pt}

% Table generated by Excel2LaTeX from sheet 'costs'
\begin{tabularx}{\textwidth}{l|ccc|ccc|cc|ccc}
\toprule
                    &
  \textbf{Interest} &
  \textbf{Lifetime} &
  \textbf{Availability} &
  \multicolumn{3}{c|}{\textbf{Overnight costs}} &
  \multicolumn{2}{c|}{\textbf{Efficiency}} &
  \multicolumn{2}{c}{\textbf{Marginal costs}} 
  \\
  
  \textbf{Technology} &
  \textbf{rates}&
  & 
  &
  \textbf{energy} &
  \textbf{charging power} &
  \textbf{discharging power} &
  \textbf{charging} &
  \textbf{discharging} &
  \textbf{charging} &
  \textbf{discharging}
  \\
             &
             &
  [years]    &
             &
  [1000 EUR] &
  [1000 EUR] &
  [1000 EUR] &
             &
             &
       [EUR] &
       [EUR] &
  \\
\midrule
Li-ion battery        & \multirow{3}{*}{0.04} & 20 & 0.98 & 142 & 80  & 80  & 0.96 & 0.96 & 0.5 & 0.5 \\
Pumped hydro          &                       & 80 & 0.89 & 10  & 550 & 550 & 0.97 & 0.91 & 0.5 & 0.5 \\
Power-to-gas-to-power &                       & 25 & 0.95 & 2   & 550 & 435 & 0.73 & 0.42 & 0.5 & 0.5 \\
\bottomrule
\end{tabularx}%

\label{tab:table1_a}
\end{subtable}

\vspace{1cm}

\begin{subtable}[t]{\textwidth}

\caption{Electricity generation}

\tiny

\centering

% Table generated by Excel2LaTeX from sheet 'costs'
\begin{tabularx}{.9\textwidth}{l|cccccccc}

    \toprule
    \textbf{Technology}      &
    \textbf{Interest rates}  &
    \textbf{Lifetime}        &
    \textbf{Availability}    &
    \textbf{Overnight costs} &
    \textbf{Fixed costs}     &
    \textbf{Efficiency}      &
    \textbf{Carbon content}  &
    \textbf{Fuel costs}
    \\
                             &
                             &
    [years]                  &
                             &
    [1000 EUR]               &
    [1000 EUR]               &
                             &
    [t/MWh]                  &
    [EUR/MWh]

    \\

    \midrule

    Run-of-river             & \multirow{12}[2]{*}{0.04} & 50 & 1.00 & 3,000 & 30  & 0.90 & 0.00 & 0    \\
    Nuclear                  &                           & 40 & 0.91 & 6,000 & 30  & 0.34 & 0.00 & 3.4  \\
    Lignite                  &                           & 35 & 0.95 & 1,500 & 30  & 0.38 & 0.40 & 5.5  \\
    Hard coal                &                           & 35 & 0.96 & 1,300 & 30  & 0.43 & 0.34 & 8.3  \\
    CCGT                     &                           & 25 & 0.96 & 800   & 20  & 0.54 & 0.20 & 30.0 \\
    OCGT                     &                           & 25 & 0.95 & 400   & 15  & 0.40 & 0.20 & 30.0 \\
    Oil                      &                           & 25 & 0.90 & 400   & 6.7 & 0.35 & 0.27 & 29.0 \\
    Other                    &                           & 30 & 0.90 & 1,500 & 30  & 0.35 & 0.35 & 18.1 \\
    Bioenergy                &                           & 30 & 1.00 & 1,951 & 100 & 0.49 & 0.00 & 32.5 \\
    Wind onshore             &                           & 25 & 1.00 & 1,182 & 35  & 1.00 & 0.00 & 0    \\
    Wind offshore            &                           & 25 & 1.00 & 2,506 & 100 & 1.00 & 0.00 & 0    \\
    Solar photovoltaic       &                           & 25 & 1.00 & 400   & 25  & 1.00 & 0.00 & 0    \\
    \bottomrule
\end{tabularx}%

\label{tab:table1_b}
\end{subtable}
\label{tab:costs}
\end{table}

\noindent Table \ref{tab:costs} provides an overview of the cost assumptions of the model.

\newpage

% -------------
\subsection{Supplementary Notes}
% -------------

\subsubsection{Baseline}

In the following, we show further results of our baseline scenario runs.

\begin{figure}[H]
\vspace{0.5cm}
    \centering
    \caption{Capacity investments under baseline assumptions}
    \includegraphics[width=\textwidth]{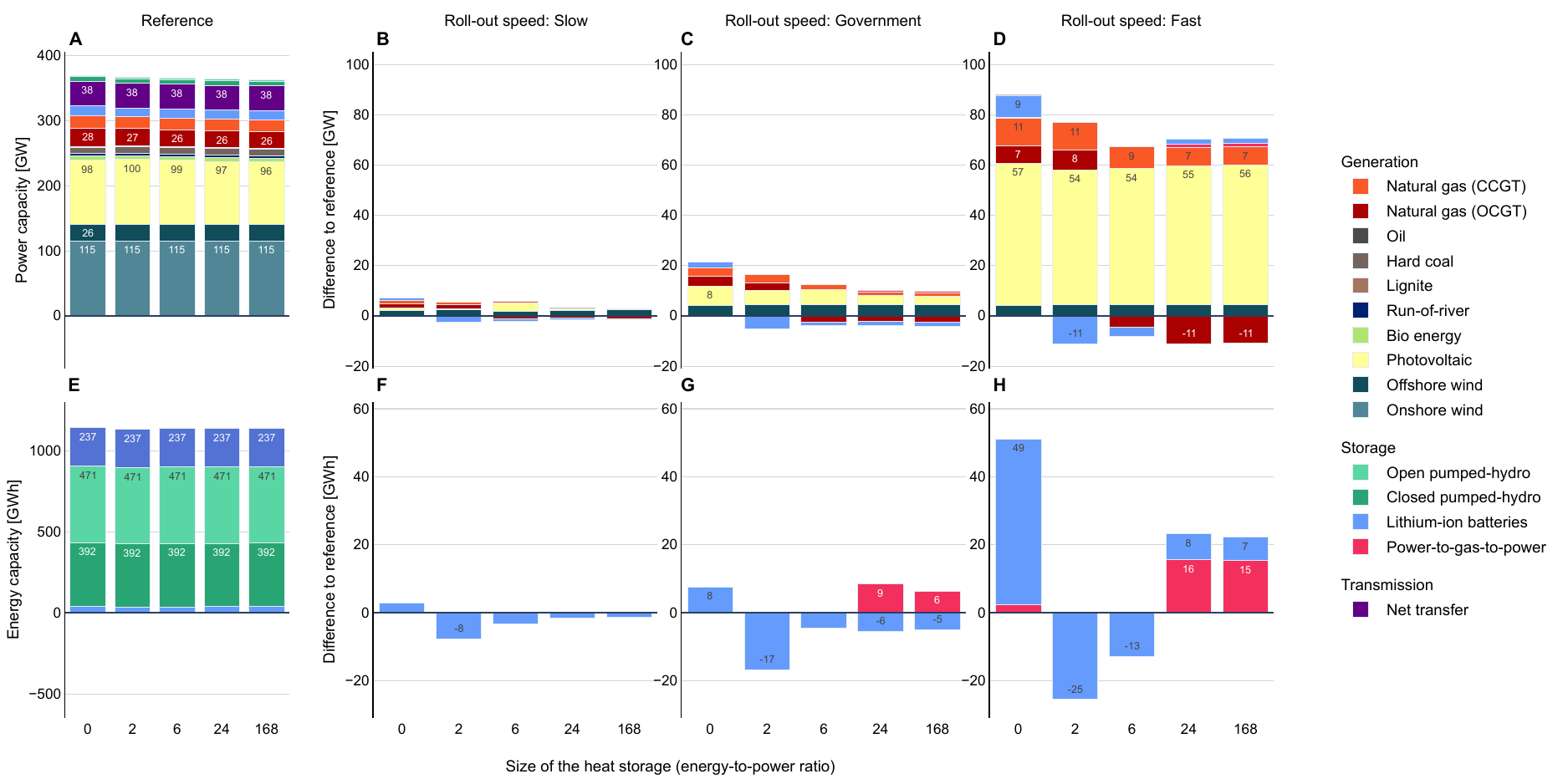}
    \begin{minipage}[c]{\textwidth}
    \bigskip
    \footnotesize
    Absolute values in the reference scenarios (\textbf{A} and \textbf{E}) and changes to the respective reference scenarios depending on the rollout scenario (\textbf{B}-\textbf{D} and \textbf{F}-\textbf{H})  and the size of heat pump storage. Values for Germany are depicted.
    \end{minipage}
    \label{fig:investments_full}
\end{figure}

Figure \ref*{fig:investments_full} shows the complete optimal capacity choices of the model. The effects of additional heat storage of heat pumps on optimal battery storage energy capacities (lower row of panels, Figure~\ref{fig:investments_full}) are pronounced. Inflexible heat pumps lead to an increased need for storage energy; battery energy storage increases by eight~GWh in the \textit{government} rollout (panel~G) by and 49~GWh in the \textit{fast} rollout (panel~H). With a two~hour heat storage, this finding reverses and the optimal energy capacity of batteries decreases by 17~GWh or 25~GWh, respectively. Beyond a two~hour heat storage, effects are smaller. The results presented in Figure~\ref{fig:investments_full} show that especially short-duration electricity storage and heat buffer storage are substitutes.

\begin{figure}[H]
\vspace{0.5cm}
    \centering
    \caption{Total electricity generation under baseline assumptions}
    \includegraphics[width=\textwidth]{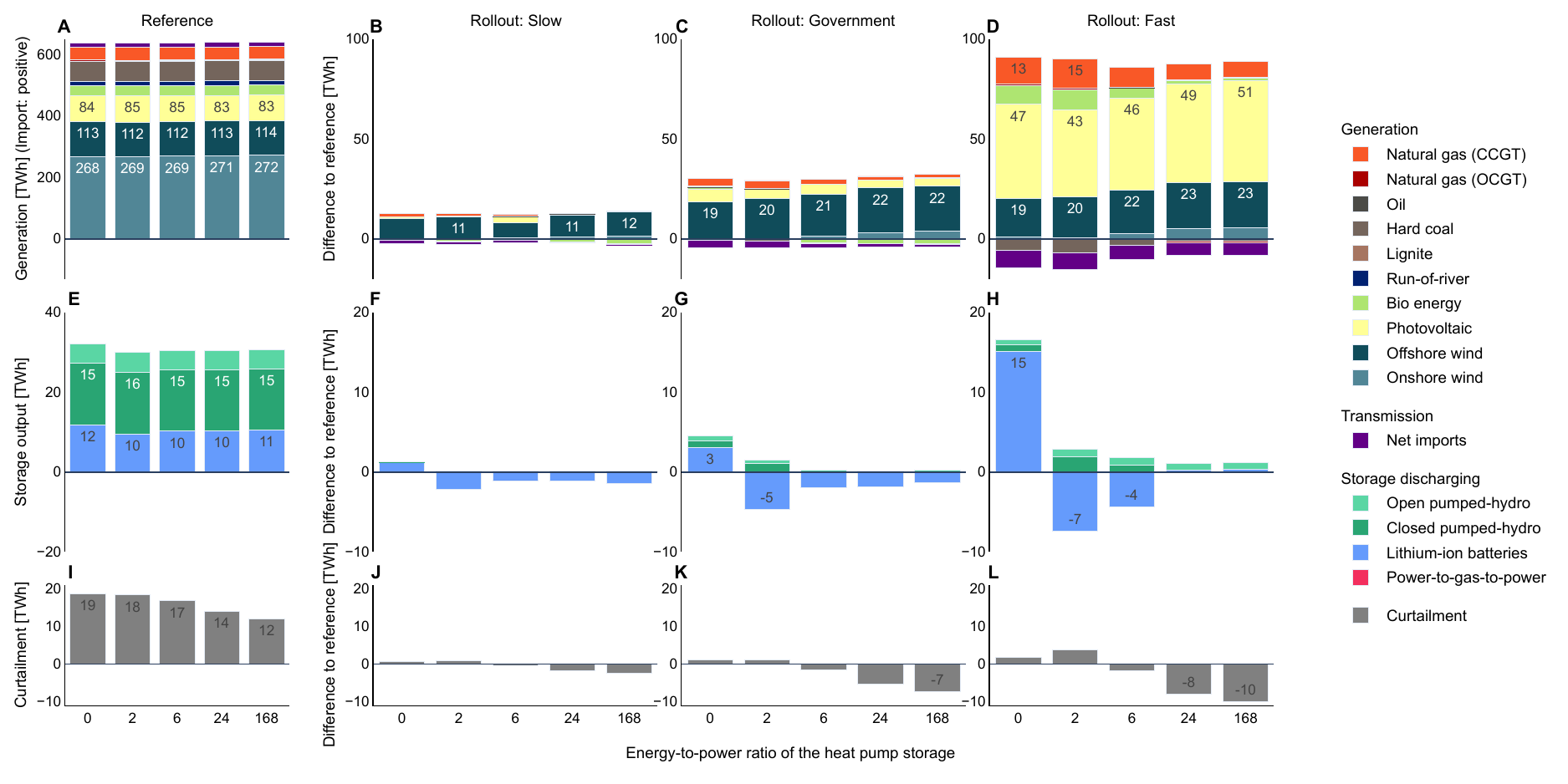}
    \begin{minipage}[c]{\textwidth}
    \bigskip
    \footnotesize
    Absolute values in the reference scenarios (\textbf{A}, \textbf{E}, and \textbf{I}) and changes to the respective reference scenarios depending on the rollout scenario (\textbf{B}-\textbf{D}, \textbf{F}-\textbf{H}, and \textbf{J}-\textbf{L}) and the size of heat pump storage. Values for Germany are depicted.
    \end{minipage}
    \label{fig:dispatch_full}
\end{figure}

Figure \ref{fig:dispatch_full} depicts the complete dispatch results from Germany, including all references and renewable curtailment.

\begin{figure}[H]
\vspace{0.5cm}
    \centering
    \caption{Capacity changes in other countries due to an introduction of heat pumps in Germany}
    \includegraphics[width=.95\textwidth]{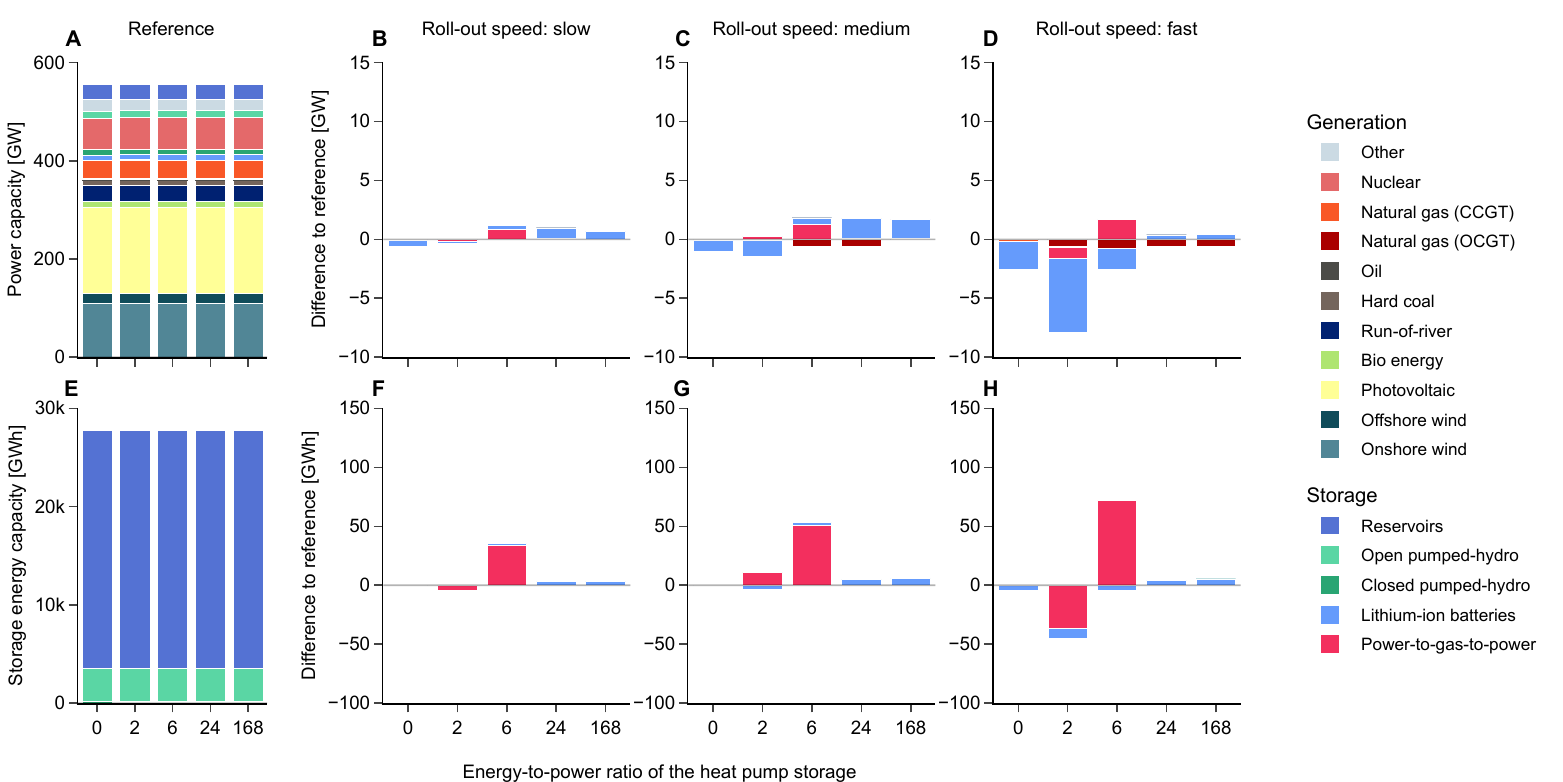}
    \begin{minipage}[c]{\textwidth}
    \bigskip
    \footnotesize
    Results for \textit{reference} scenarios (\textbf{A} and \textbf{E}).  Respective differences to the reference (\textbf{B}-\textbf{D} and \textbf{F}-\textbf{H}). Results are shown for scenarios under \textit{baseline}.
    \end{minipage}
    \label{fig:investment_other_countries}
\end{figure}

Figure \ref{fig:investment_other_countries} depicts the aggregated capacities and other countries and how they change with a heat pump rollout in Germany. The introduction of heat pumps in Germany does not affect the aggregated power plant and storage portfolio of other countries on a large scale. In terms of power (discharge) capacities (upper row of Figure~\ref{fig:investment_other_countries}), the introduction leads only to minor changes, even to reductions in battery discharge capacities for some heat storage sizes. Regarding storage energy, we see capacity decreases and increases, especially for long-duration storage; however, there is no consistent pattern that would mean that the expansion of heat pumps in Germany is mainly supported by generation and storage capacities outside Germany. 

\begin{figure}[H]
\vspace{0.5cm}
\centering
\caption{Heat output, heat pump electric load with different storage sizes, and residual load (\textit{fast} rollout)}
\includegraphics[width=\textwidth]{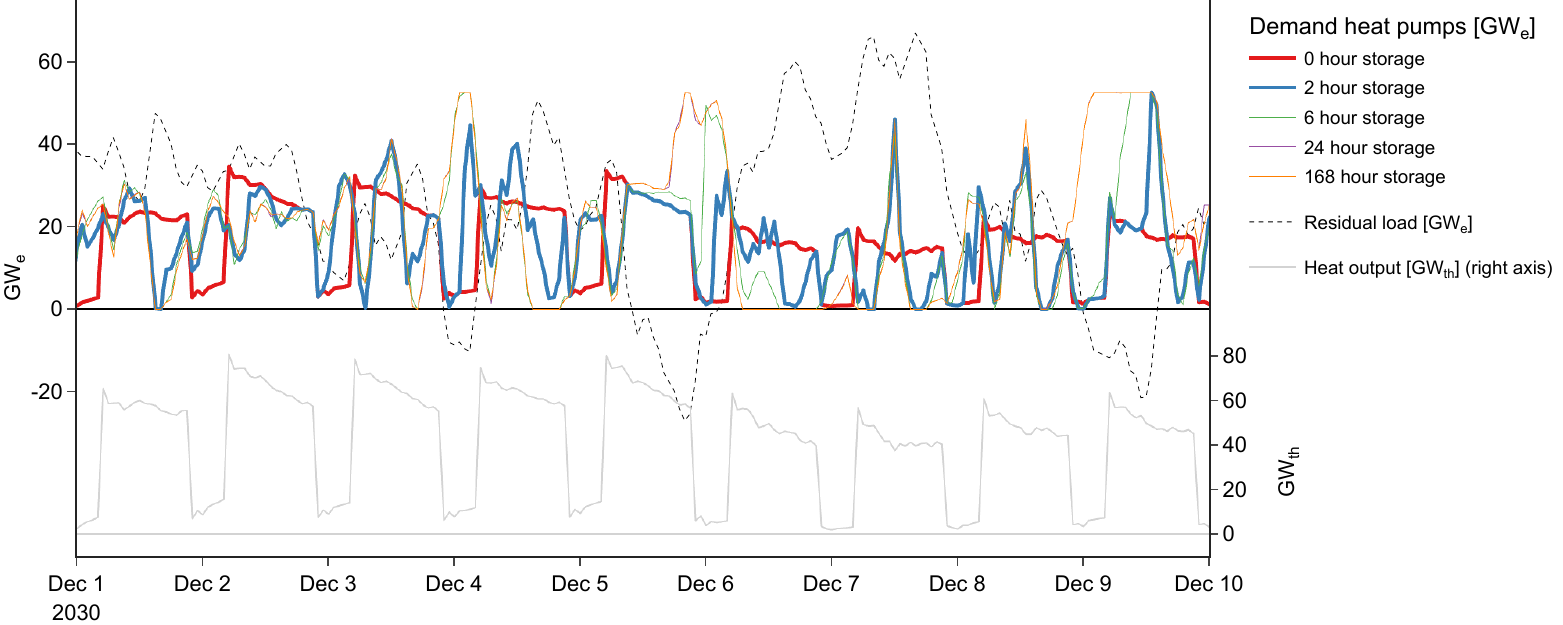}
\begin{minipage}[c]{0.97\textwidth}
    \bigskip
    \footnotesize
    Residual load, the heat output of heat pumps, and their electricity demand for different heat storage sizes are shown for \textit{baseline} assumption in a \textit{fast} rollout.
\end{minipage}
\label{fig:hp-draw-fast}
\end{figure}

Figure \ref{fig:hp-draw-fast} shows the profiles of heat output and electricity demand of heat pumps (for different heat storage sizes) and the residual load. The decoupling of heat output/demand and electricity consumption of heat pumps can be seen easily, even for small heat storage sizes. The heat pumps aim to align their consumption to the hours of lowest residual demand, i.e.,~often around midday with high generation of electricity from PV.

\begin{figure}[H]
\vspace{0.5cm}
    \centering
    \caption{Residual load duration curves (first 50 hours)}
    \includegraphics[width=\textwidth]{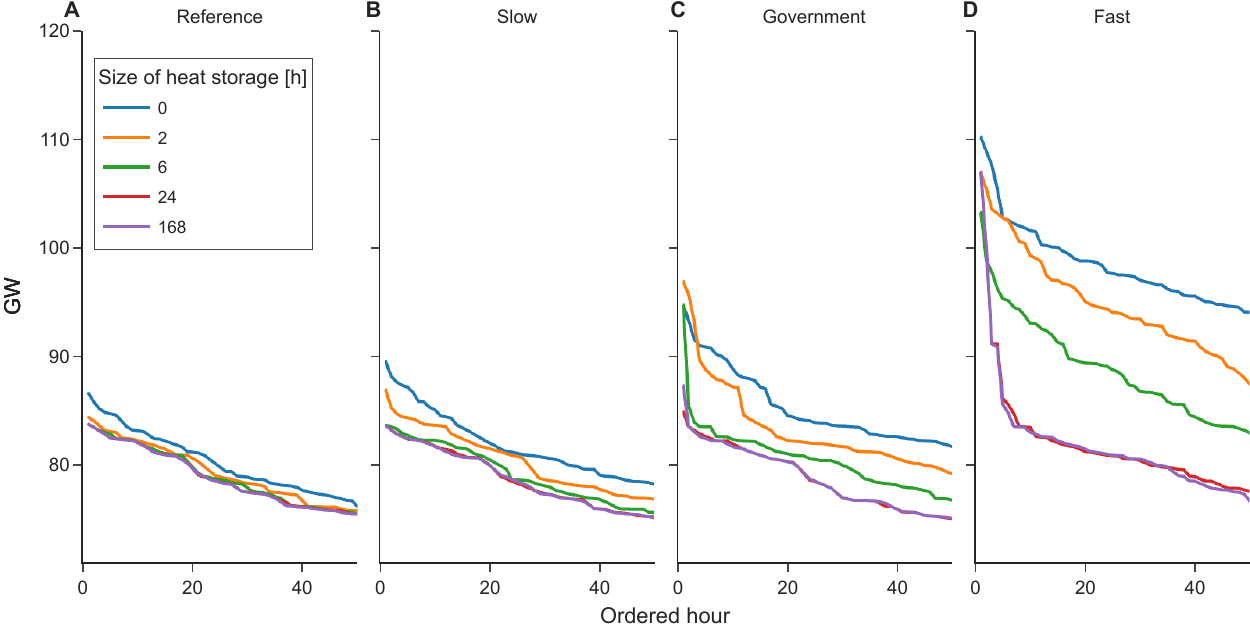}
    \begin{minipage}[c]{0.95\textwidth}
    \bigskip
    \footnotesize
    \textbf{A}-\textbf{D} Residual load duration curves in Germany are shown for different rollout speeds of heat pumps (columns) and their heat storage size (color). A residual load duration curve depicts hourly residual loads (load minus generation of renewable energies), sorted in descending order.
    \end{minipage}
    \label{fig:rldc}
\end{figure}

Figure \ref{fig:rldc} shows the impact of additional heat pumps on residual load duration curves and how heat storage changes these. The peak load-increasing effects of heat pumps can be clearly seen moving from the \textit{reference} rollout to more ambitious rollouts. The peak load-reducing effect of heat storage is especially pronounced in the \textit{government} and \textit{fast} rollout scenarios.

Table \ref{tab:savings_fullresults} provides an overview of the full results of natural gas and emissions savings.

% Table generated by Excel2LaTeX from sheet 'summary results'
\begin{table}[H]
  \centering
  \caption{Full results of natural gas and emissions savings}
%    \toprule
    \resizebox{\textwidth}{!}{
    \begin{tabular}{p{18.715em}lrrrrrr}
    \hline
    Gas price & Euro  & \multicolumn{2}{c}{50 Euro} & \multicolumn{2}{c}{100 Euro} & \multicolumn{2}{c}{150 Euro} \\
    Heat pump rollout &       & \multicolumn{1}{c}{Gov.} & \multicolumn{1}{c}{Fast} & \multicolumn{1}{c}{Gov.} & \multicolumn{1}{c}{Fast} & \multicolumn{1}{c}{Gov.} & \multicolumn{1}{c}{Fast} \\
    \hline
    Total gas displaced by heat pumps & TWh\textsubscript{th}        & -103.21	& -251.43	& -103.21	& -251.43	& -103.21 & -251.43 \\
    Change in  gas displaced by heat pumps & TWh\textsubscript{th}   & -75.75	& -223.97	& -75.75	& -223.97	& -75.75  & -223.97 \\
    \multicolumn{1}{l}{Total electricity produced from gas} & TWh    & +219.59	& +227.87	& +76.88	& +82.86	& +70.72  &	+78.97 \\
    Change in electricity produced from gas & TWh                    & +1.33	& +9.61	    & +1.00	    & +6.97	    & +0.00	  & +10.81 \\
    Total gas use for electricity generation & TWh\textsubscript{th} & +407.44	& +422.86	& +145.27	& +156.10	& +131.44 & +146.89 \\
    Change in gas use for electricity & TWh\textsubscript{th}        & +2.58	& +18.01	& +1.92	    & +12.75	& +4.90	  & +20.35 \\
    Gas savings & TWh\textsubscript{th}                              & -73.17	& -205.96	& -73.83	& -211.22	& -70.85  & -203.62 \\
    Emission savings & Mio tCO2eq                                    & -14.63	& -41.19	& -14.77	& -42.24	& -14.17  & -40.72 \\
    Cost savings & billion Euro                                      & -2.05	& -6.73	    & -5.48	    & -16.80	& -9.07	  & -27.07 \\
    \hline
    \end{tabular}
    }
    \begin{minipage}[c]{\textwidth}
        \bigskip
        \footnotesize
        The changes shown in the table are calculated with respect to the corresponding reference scenario (with the same natural gas price).
    \end{minipage}
  \label{tab:savings_fullresults}%
\end{table}%

% ---------------------------------------------------------
\subsubsection{Sensitivity analyses} \label{sec:sensitivity-appendix}
% ---------------------------------------------------------

In addition to our baseline scenario runs in which we vary the rollout speed of heat pumps and heat storage duration, we conduct several sensitivity analyses. Those help us judge how strongly our results hinge on certain fundamental model assumptions. Table~\ref{tab:sensitivity scenarios} provides an overview of all sensitivity analyses conducted.

% Table generated by Excel2LaTeX from sheet 'scenarios'
\begin{table}[H]
    \centering
    \small
    \caption{Overview of sensitivity analyses}
    \begin{tabular}{r|l|l}
          \toprule
          & \textbf{Scenario acronym}            & \textbf{Description}                                                                   \\
        \midrule
        1 & \textit{no wind cap}                 & No upper bound on on- and offshore wind capacities in Germany.                         \\
        2 & \textit{gas100}                      & Natural gas price at 100~Euro~per~MWh.                                                 \\
        3 & \textit{gas150}                      & Natural gas price at 150~Euro~per~MWh.                                                 \\
        4 & \textit{coal phase-out}              & No coal-fired plants allowed in Germany.                                               \\
        5 & \textit{coal phase-out + gas100}     & Combination of 2 and 4.                                                                \\
        6 & \textit{coal phase-out + gas150}     & Combination of 3 and 4.                                                                \\
        7 & \textit{RE drought}                  & All renewable energy capacity factors of all countries set to zero in one winter week. \\
        8 & \textit{RE drought + coal phase-out} & Combination of 4 and 7.                                                                \\
        \bottomrule
    \end{tabular}%
    \label{tab:sensitivity scenarios}%
\end{table}%

\begin{figure}[H]
\vspace{0.5cm}
    \centering
    \caption{Capacity investments in different sensitivity analyses}
    \includegraphics[width=\textwidth]{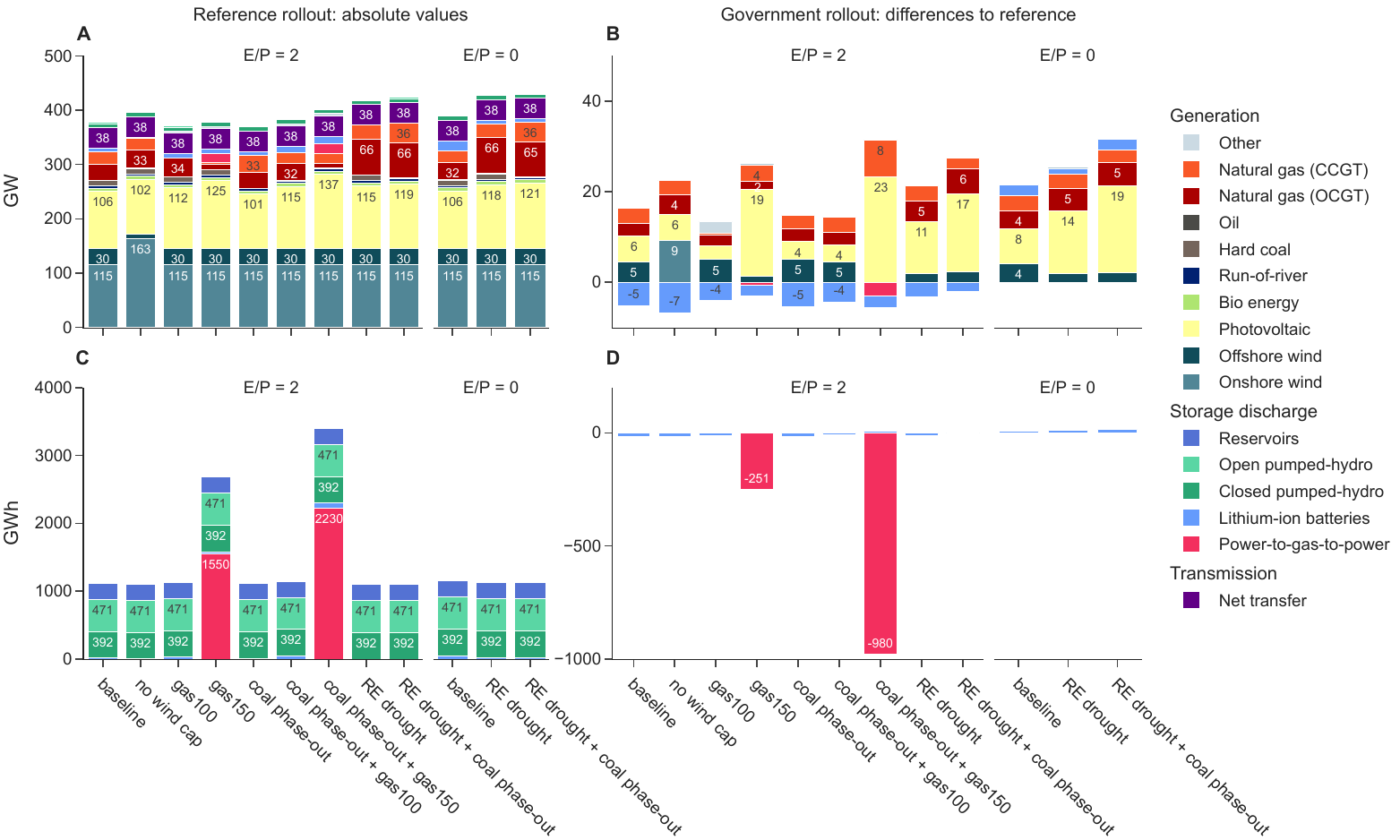}
    \begin{minipage}[c]{0.95\textwidth}
        \bigskip
        \footnotesize
        \textbf{A} and \textbf{C} Optimal capacities in the \textit{reference} rollout for different scenario assumptions in the case of flexible heat pumps (with two-hour storage) and inflexible heat pumps (with no storage). \textbf{B} and \textbf{D} Changes to the respective reference for a \textit{government} rollout.\textbf{C} and \textbf{D} Respective results for storage energy. Please note the different y-axis ranges of the different panels.
    \end{minipage}
    \label{fig:investments_sens_si}
\end{figure}

\begin{figure}[H]
\vspace{0.5cm}
    \centering
    \caption{Additional power system costs of heating energy provided in different sensitivity analyses}
    \includegraphics[width=\textwidth]{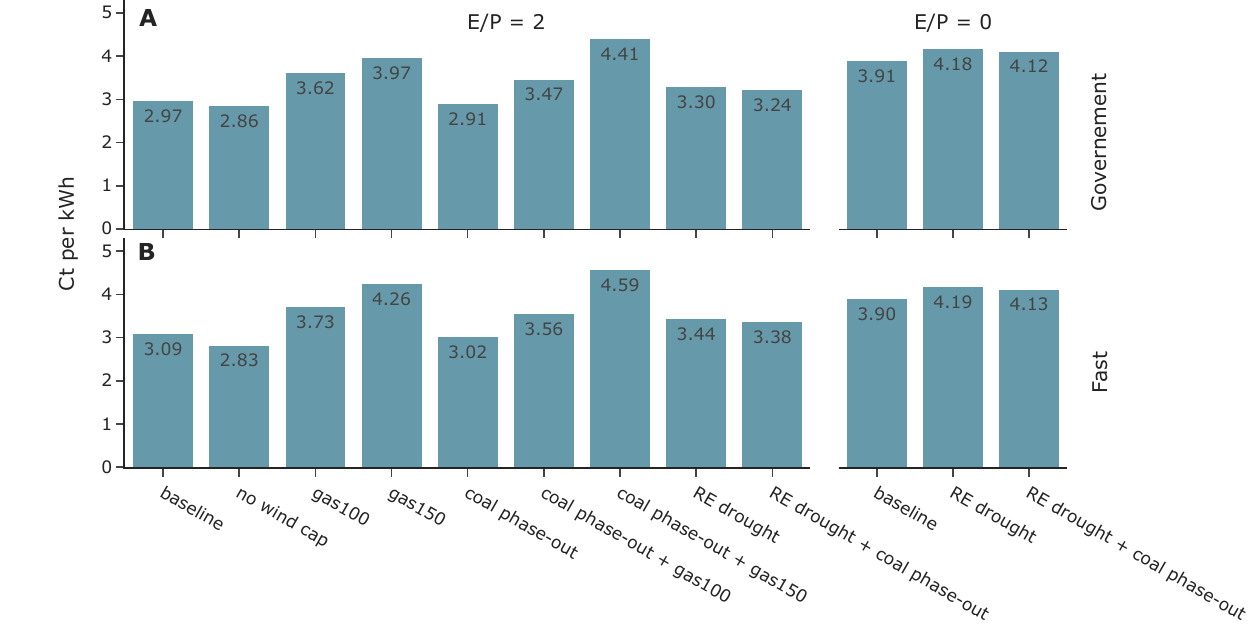}
    \begin{minipage}[c]{0.90\textwidth}
    \bigskip
    \footnotesize
    Additional system costs per heating energy provided (in Euro-cent/kilowatt-hour) in different sensitivity scenarios for a \textit{government} (\textbf{A)} and \textit{fast} rollout (\textbf{B}) with heat storage sizes of zero and two hours.
    \end{minipage}
    \label{fig:aschp_sens_si}
\end{figure}

\begin{figure}[H]
\vspace{0.5cm}
    \centering
    \caption{Yearly electricity generation in different sensitivity analyses}
    \includegraphics[width=\textwidth]{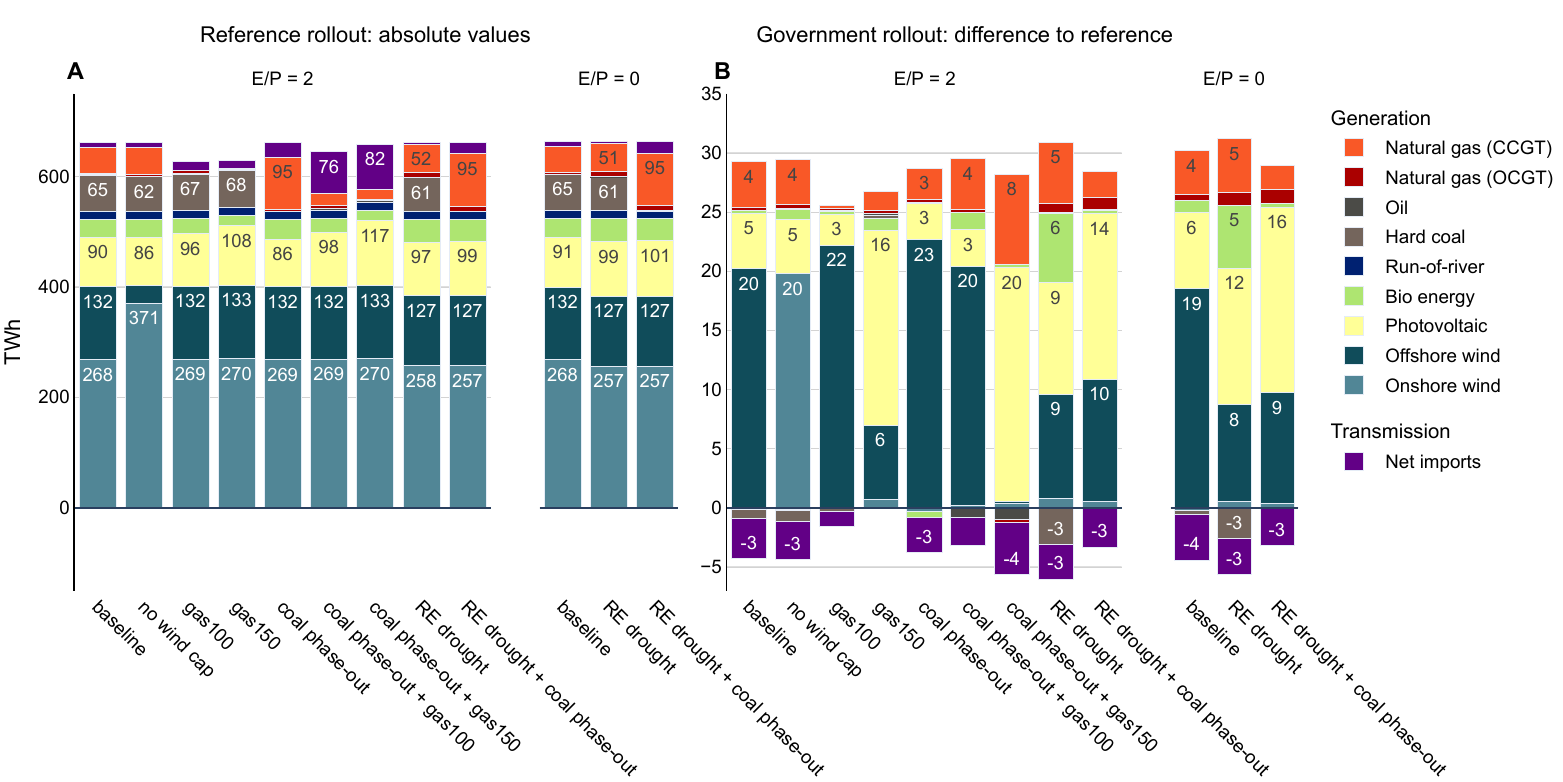}
        \begin{minipage}[c]{0.90\textwidth}
        \bigskip
        \footnotesize
        \textbf{A} Optimal dispatch in the \textit{reference} rollout for different scenario assumptions in case of a two-hour and zero-hour heat storage. \textbf{B} Changes to the respective reference for a \textit{government} rollout. Please note the different y-axis ranges of the different panels.
        \end{minipage}
    \label{fig:dispatch_sens}
\end{figure}

In the following, we briefly present the different sensitivity analyses and discuss their results in terms of capacity investments (Figure~\ref{fig:investments_sens_si}), dispatch (Figure~\ref{fig:dispatch_sens}), and additional system costs of heating provided (Figure~\ref{fig:aschp_sens_si}).

% -----
\paragraph{No capacity expansion limit of wind energy (\textit{no wind cap})}
% -----

In the baseline scenarios, we assume an upper limit for on- and offshore wind power capacity expansion in Germany of 115~GW and 30~GW, respectively. This appears to be realistic and policy-relevant for 2030. Assuming unbounded wind power capacity expansion, considering real-world constraints related to regulation, land availability, and public acceptance, seems not feasible. However, in a sensitivity analysis, we drop this upper limit so that expansion of on- and offshore wind power capacities are unconstrained, and we assess what a less constrained optimal solution looks like.

The removal of the upper bound for wind power leads to overall higher wind capacities and slightly lower PV capacity expansion, already in the reference rollout scenario (Figure \ref{fig:investments_sens_si}, panel A). In particular, offshore wind is substituted by onshore wind due to lower costs. These changes correspond with a higher yearly generation of onshore wind energy in the reference rollout scenario (Figure~\ref{fig:dispatch_sens}, panel A) compared to the baseline scenario. Given this reference, an additional rollout of heat pumps leads to a substantial expansion of wind onshore capacities, yet far fewer additional PV capacities (Figure~\ref{fig:investments_sens_si}, panel B) than in the baseline. In consequence, additional dispatch consists mainly of onshore wind energy instead of offshore (Figure~\ref{fig:dispatch_sens}, panel B). These results confirm that the availability of wind power aligns well with the seasonality of the heating demand. Optimal storage energy installation rarely changes in comparison to the baseline (Figure~\ref{fig:investments_sens_si}, panels C and D). Despite the relatively large shift between different generation technologies, additional system costs do not change much compared to the baseline setting (Figure~\ref{fig:aschp_sens_si}). This implies that a rollout of heat pumps can also be combined with different solar PV capacity expansions in case of binding wind power capacity limits with little additional costs. %, making use of the flexibility provided by the European interconnection.

% -----
\paragraph{Sustained high gas prices (\textit{gas100} and \textit{gas150})}
% -----

Due to the Russian invasion of Ukraine, the natural gas supply structure of Europe was fundamentally changed. In the foreseeable future, Germany will not import any more Russian gas but will rely on more costly imports of liquefied natural gas (LNG) from other regions. Although wholesale gas prices have been falling strongly since their peak levels of over 300~Euro~per~MWh in August 2022 and range by the time of writing between 30-40~Euro~per~MWh, it remains possible that new spikes could arise in the future. In our set of baseline scenarios, we assume a natural gas price of 50~Euro per MWh. We introduce two alternative scenarios, \textit{gas100} and \textit{gas150}, in which we assume natural gas prices of 100 or 150~Euro~per~MWh.

Higher gas prices barely alter the optimal capacity expansion in the reference rollout. Even a fast heat pump rollout leads to very similar capacity installations compared to baseline assumptions, with increased solar PV and reduced gas power plant capacities for a gas price of 150~Euro~per~MWh (Figure \ref{fig:investments_sens_si}, panel B). For the \textit{gas150} scenario, we see that a considerable amount of long-duration storage is deployed, serving as a substitute. Introducing heat pumps reduces slightly these capacities. Regarding yearly energy generation, higher gas prices lead, not surprisingly, to lower electricity generation by gas-fired power plants - in the \textit{reference} rollout but also in more ambitious rollouts. Noteworthy is that the additional electricity generated due to the heat pumps comes in the scenario \textit{gas150} mainly from PV, as already the capacity results suggest. This outcome can be explained by the capacity limits of offshore wind power that is already reached in the \textit{reference} rollout. Additional power system costs per heating unit increase substantially compared to the baseline because of more expensive natural gas \ref{fig:aschp_sens_si}, suggesting that the model cannot fully substitute gas-fired power plants. Overall, we do not observe substantial changes compared to our scenarios under \textit{baseline} assumptions with a lower gas price.

% -----
\paragraph{Coal phase-out (\textit{coal phase-out})}
% -----

In the baseline scenarios, we allow coal-fired power plants to generate electricity in 2030, in accordance with the currently planned German coal phase-out by 2038. However, the current governmental coalition agreed to ``ideally bring forward'' the coal phase-out to 2030. Although this agreement has not yet been translated into binding law, we aim to analyze the power sector consequences of an earlier coal phase-out combined with a faster heat pump rollout.

In the reference rollout, coal-fired power plants that are present in the baseline scenario are mainly replaced by gas-fired plants (Figure \ref{fig:investments_sens_si}, panel A), but hardly differ otherwise. The additional capacities due to heat pump rollout are very similar to our \textit{baseline} scenarios (Figure \ref{fig:investments_sens_si}, panel B). In terms of dispatch, generation by coal-fired power plants in the \textit{reference} rollout is mainly compensated by gas-fired (CCGT) plants, as well as by increased net imports. Expanding heat pumps leads to largely similar dispatch effects as in the baseline (Figure \ref{fig:dispatch_sens}). Additional power system costs due to heat pumps are very similar to the baseline (Figure \ref{fig:aschp_sens_si}).

We also combine the coal phase-out with higher gas prices (scenarios \textit{coal phase-out + gas100} and \textit{coal phase-out + gas150}). In consequence, we see slightly higher solar PV capacity installations in the \textit{reference} rollout. Additional capacities in the \textit{government} rollout are similar to those of \textit{coal phase-out} and \textit{gas150}: either they are very similar to the \textit{baseline} scenarios, or more PV capacities are triggered in case of high gas prices (\textit{coal phase-out + gas150}). Like in scenario \textit{gas150}, offshore wind power is already at its upper bound in scenario \textit{coal phase-out + gas150}; hence, additional capacities are mainly PV. Like in \textit{gas150}, long-duration storage is installed. In terms of dispatch, results do not differ too much from the baseline, either. For the reference rollout, the missing coal-fired generation is partly displaced by electricity net imports. Yet, these net imports diminish with additional heat pumps in the fast rollout. For scenario \textit{coal phase-out + gas150}, following the capacity results, additional generation is mainly PV. Overall, additional dispatch does not vary strongly between these sensitivity scenarios and the baseline. Yet, the combination of a coal phase-out and higher gas prices leads to considerably higher power system costs because of the higher production costs of gas-fired power plants.

% -----
\paragraph{A week of a renewable energy drought (\textit{RE drought})}
% -----

As the share of variable renewable energy increases, the security of supply during prolonged periods with low renewable energy supply becomes an increasing concern. To simulate an extreme case of such a week, we artificially set wind and solar PV capacity factors to zero in all modeled countries during one winter week (scenario \textit{RE drought}). As a consequence, substantially more gas-fired power plants are installed in the \textit{reference} (Figure \ref{fig:investments_sens_si}, panel~A). Adding more heat pumps does not change the optimal capacities much compared to the \textit{baseline} setting. We can see a capacity-reducing effects of heat storage, when comparing the results of the scenarios with flexible and inflexible heat pumps. The results regarding generation (Figure~\ref{fig:dispatch_sens}) also do not differ much compared to the \textit{baseline}. Additional system costs per heat provided (Figure~\ref{fig:aschp_sens_si}) are increased, and heat storage also reduces power sector costs in this scenario. Combining this sensitivity with a coal phase-out (scenario \textit{RE drought + coal phase-out}) does not alter the results substantially. The principal difference is that coal-fired power plants are replaced by gas-fired power plants. Note that long-duration electricity storage does not play much of a role in the renewable energy drought modeled here, as it would be more expensive than providing a backup capacity of OCGT plants. This would change in case of substantially higher natural gas prices (compare sensitivity \textit{gas150}), or if the potential to build new gas-fired power plants was restricted, or if long-duration electricity storage became substantially cheaper than assumed here.

\begin{figure}[H]
\vspace{0.5cm}
\centering
    \caption{Heat output, heat pump electric load with different storage sizes, and residual load (\textit{RE drought)}}
    \includegraphics[width=\textwidth]{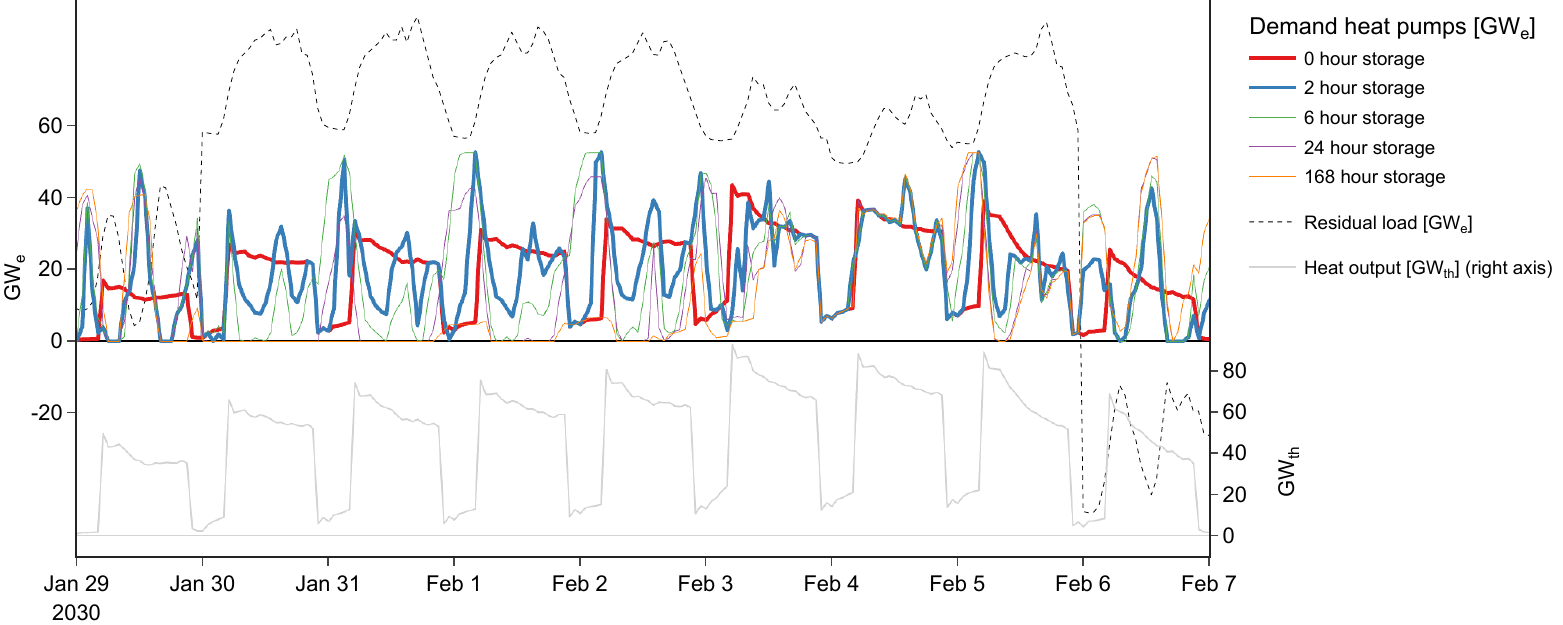}
    \begin{minipage}[c]{\textwidth}
    \bigskip
    \footnotesize
    Residual load, the heat output of heat pumps, and their electricity demand for different heat storage sizes in the \textit{RE drought} scenario with a \textit{fast} rollout.
    \end{minipage}
    \label{fig:hp-draw-fast-badweek}
\end{figure}

Figure \ref{fig:hp-draw-fast-badweek} shows the electricity consumption pattern in a week of a renewable energy drought. Despite a consistent positive residual load, we see stark differences in heat pump electricity demand depending on their heat storage size. Comparable to a regular week, heat pumps shift their consumption to the periods with the lowest residual load. That is even done for the smallest heat storage size to the largest extent possible.

% -------------
\end{document}